\documentclass{birkjour}
\usepackage[applemac]{inputenc}
\usepackage[cyr]{aeguill}
\usepackage[pdftex]{thumbpdf,hyperref}
\usepackage{amssymb}
\usepackage{tikz}
\usepackage{graphicx}
\usepackage{longtable}
\usepackage{subfigure}
\usepackage{enumerate}

\usepackage[all,knot,poly]{xy}
\usepackage{youngtab,mathrsfs,url}

\newcommand{\e}{\mathrm{e}}
\newtheorem{example}{Example}

\newcommand{\Sym}{\mathcal S}

\renewcommand{\L}{\mathscr{L}}
\newcommand{\Knot}{\mathcal K}
\renewcommand{\bar}[1]{\overline{#1}}
\renewcommand{\hat}[1]{\widehat{#1}}

\newcommand{\intint}[1]{\left[\kern-0.15em\left[#1\right]\kern-0.15em\right]}
\newcommand{\R}{\mathbb{R}}
\newcommand{\C}{\mathbb{C}}
\newcommand{\N}{\mathbb{N}}
\newcommand{\ZZ}{\mathbb{Z}}
\newcommand{\Q}{\mathbb{Q}}

\renewcommand{\vec}[1]{\mathbf{#1}}

\newcommand{\abs}[1]{\vert#1 \vert}

\newcommand{\bigsqplus}{\qed\kern-0.75em+} 
\newcommand{\sqplus}{\qed\kern-1em+}
\newcommand{\semsum}{\odot\kern-0.75em\ltimes}

\newcommand{\tens}[1]{\mathbf{#1}}

\DeclareMathOperator{\Tr}{\mathrm{Tr}}

\DeclareMathOperator{\Res}{\mathrm{Res}}

\def\XXint#1#2#3{{\setbox0=\hbox{$#1{#2#3}{\int}$}\vcenter{\hbox{$#2#3$}}\kern-.5\wd0}}
\newcommand{\UD}[1]{\mathcal{D}#1}
\newcommand{\ud}[1]{\mathrm{d}#1}

\newcommand{\bra}[1]{\langle #1|}
\newcommand{\ket}[1]{|#1\rangle}
\newcommand{\av}[1]{\langle #1\rangle}

\newcommand{\Part}{\mathscr{P}}

\renewcommand{\S}{\mathbb{S}}
\newcommand{\D}{\mathbb{D}}
\newcommand{\Torus}{\mathbb{T}}

\renewcommand{\O}{\mathcal{O}}

\renewcommand{\sl}{\mathfrak{sl}}

\newcommand{\g}{\mathfrak{g}}

\newcommand{\W}{\mathcal{W}}

\newcommand{\Hilb}{\mathcal{H}}
\newcommand{\AmSLaTeX}{{\protect\the\textfont2 A}%
\kern-.1667em\lower.5ex\hbox {\protect\the\textfont2 M}%
\kern-.125em{\protect\the\textfont2 S}-\LaTeX}
\renewcommand{\labelenumi}{$(\roman{enumi})$}

\title{Torus Knots in Lens Spaces \& Topological Strings}
\author{Sebastien Stevan}
\address{Universit\'e de Gen\`eve\\
Section de math\'ematiques\\
Case postale 64\\
1211 Gen\`eve 4 (Switzerland)}
\email{sebastien.stevan@gmail.com}
\thanks{\textbf{Acknowledgments.} We are grateful to Marcos Mari\~no for his support and for comments on the manuscript. We also thank the anonymous referee for his or her suggestions. We finally thank Van Quach Hongler for discussions on knots in lens spaces.}

\begin{document}
\begin{abstract}
We study the invariant of knots in lens spaces defined from quantum Chern-Simons theory. By means of the knot operator formalism, we derive a generalization of the Rosso-Jones formula for torus knots in $L(p,1)$. In the second part of the paper, we propose a B-model topological string theory description of torus knots in $L(2,1)$.
\end{abstract}
\maketitle
\section{Introduction}
The HOMFLY polynomial \cite{Freyd85a} is an invariant of links in the $3$-sphere, defined by a linear skein relation involving planar projections. It can be extended to rational homology spheres which are either atoroidal or Seifert-fibered; in this case, Kalfagianni \& Lin \cite{Kalfagianni99a} showed the existence and uniqueness of a power series invariant satisfying the HOMFLY skein relation. 

Apart from the $3$-sphere $\S^3$, detailed studies occur for solid tori \cite{Gabai90a,Gabai89a,Berge91a} and lens spaces $L(p,q)$. For knots in the projective space $\R P^3$, Drobotukhina \cite{Drobotukhina91a} has generalized the Reidemeister moves and the Jones polynomial. The twisted Alexander polynomial, introduced by Lin \cite{Lin01a} for links in $\S^3$, has been studied by Huynh \& L\^e \cite{Huynh08a} in $\R P^3$. Independently, using the skein module approach, Hoste \& Przytycki have extended the Jones polynomial to knots in $L(p,q)$ \cite{Przytycki91a,Hoste93a}.

Recently, Cattabriga \textit{et al.} \cite{Cattabriga12a} have generalized the Reidemeister moves to lens spaces and obtained results on the fundamental group of the link complement and the twisted Alexander polynomial. Furthermore, using a different approach based on grid diagrams, Cornwell has proved that the invariant satisfying the HOMFLY skein relation is a polynomial \cite{Cornwell10a}.

The colored generalization of the HOMFLY polynomial, defined from HOMFLY skein theory \cite{Morton93a} or from representations of quantum groups, can in principle be generalized to $3$-manifolds other than $\S^3$. The approach based on representations of quantum groups is hard to extend to general $3$-manifolds, as it involves plane projections. Even in $\S^3$, the computation of quantum invariants is anyway very challenging and explicit formulae exist only for a small class of knots.

Along with the mathematical research on links, the surprising relationship between link invariants and quantum Chern-Simons (CS) theory unveiled by Witten \cite{Witten89a} provides a very different point of view on knot theory. By means of infinite-dimensional functional integrals, CS theory gives an intrinsically three-dimensional, albeit non-rigorous definition of the colored HOMFLY polynomial. Furthermore, the construction naturally encompasses different quantum group invariants in various $3$-manifolds. In particular, we will make the point that torus knots in lens spaces can be successfully analyzed within this framework.

An even richer perspective on topological invariants coming from physics stems from the relationship between CS theory and topological string theory. The duality conjecture by Gopakumar \& Vafa \cite{Gopakumar99a} has numerous implications in knot theory \cite{Ooguri00a}. The A-model description provides enumerative properties of knot invariants whereas, by mirror symmetry, the B-model formalism conjecturally associates to each knot an algebraic curve. Recently, Brini \textit{et al.} \cite{Brini11a} have extended this description to torus knots in $\S^3$. Furthermore, CS theory on $L(p,1)$ also possesses a topological string theory description \cite{Aganagic04a}. As an application of the B-model remodeling \cite{Bouchard09a}, this description has been explicitly worked out for the unknot. Adapting the method of Brini \textit{et al.}, we will derive an algebraic curve describing torus knots in $\R P^3$.

Let us conclude this introduction by a short outline of the paper. In Section \ref{knot_inv}, we recall a few generalities on knot invariants and lens spaces, and introduce CS theory and the knot operator of Labastida \textit{et al.} \cite{Labastida89a,Labastida91a}. We compute invariants of torus knots in lens spaces in Section \ref{cs_lens}. In Section \ref{b-modelS3}, we review the B-model description of torus knots in $\S^3$. The main result of this paper is presented in Section \ref{b-modelRP3}, where we propose an algebraic curve describing torus knots in $\R P^3$.

\section{Invariants of Knots}\label{knot_inv}
We begin by defining the invariants of knots in the $3$-sphere and in other $3$-manifolds. We focus in particular on torus knots in lens spaces. In section \ref{tqft}, we recall the knot invariants constructed from quantum Chern-Simons theory, and the knot operator formalism.

\subsection{Knots in the $3$-Sphere}
We call knot (in $\S^3$) any closed connected one-dimensional manifold $\Knot$ embedded in $\S^3$. Knot invariants are defined by means of plane projections with crossing prescriptions, that are called knot diagrams. The (framed) HOMFLY skein of the annulus, originally introduced by Turaev \cite{Turaev90a}, is defined as the set of linear combinations with coefficients in $\ZZ[t^{\pm 1},v^{\pm 1}]$ of diagrams in the annulus $\S^1\times[0,1]$, quotiented by the skein relation
\begin{center}
\begin{tikzpicture}
\draw[shift={(-3 cm,0 cm)}] [thick] (1.5,0) -- (0.85,0.65);
\draw[shift={(-3 cm,0 cm)}]  [->,thick] (0.65,0.85) -- (0,1.5);
\draw[shift={(-3 cm,0 cm)}] [->,thick] (0,0) -- (1.5,1.5);
\draw (-0.75,0.7) node {$-$};
\draw[shift={(-3 cm,0cm)},dashed] (0.75,0.75) ellipse (1.41*0.75 and 1.41*0.75);
\draw[shift={(-3 cm,0 cm)}] (1,-0.5) node [below] {$L_+$};
\draw [thick] (0,0) -- (0.65,0.65);
\draw  [thick,->] (0.85,0.85) -- (1.5,1.5);
\draw [thick,<-] (0,1.5) -- (1.5,0);
\draw (3.25,0.75) node {$=\quad(t-t^{-1})$};
\draw[dashed] (0.75,0.75) ellipse (1.41*0.75 and 1.41*0.75);
\draw (0.75,-0.5) node [below] {$L_-$};
\draw[shift={(5 cm,0 cm)}]  [->, thick] (0,0) to[bend right] (0,1.5);
\draw[shift={(5 cm,0 cm)}] [->,thick] (1.5,0) to[bend left] (1.5,1.5);
\draw[shift={(5 cm,0cm)},dashed] (0.75,0.75) ellipse (1.41*0.75 and 1.41*0.75);
\draw[shift={(5 cm,0 cm)}] (1,-0.5) node [below] {$L_0$};
\end{tikzpicture}
\end{center}
and the additional framing relations
\begin{center}
\begin{tikzpicture}
\draw[thick]  (0,-1.2) .. controls (0.1,0.6) and (0.8,0.7) .. (1,0) ;
\draw[thick]  (1,0) .. controls (1.1,-0.6) and (0.5,-0.8) .. (0.3,-0.1);
\draw[thick,->]  (0.2,0.2) .. controls (0,1) .. (0,1) ;
\draw (2.5,0) node {$=\qquad v$};
\draw[thick,->,shift={(3.5 cm,0cm)}]  (0,-1.2) .. controls (0.2,0) and (-0.2,0) ..  (0,1) ;

\draw[thick,shift={(6 cm,0cm)},yscale=-1,<-] (0,-1.2) .. controls (0.1,0.6) and (0.8,0.7) .. (1,0) ;
\draw[thick,shift={(6 cm,0cm)},yscale=-1] (1,0) .. controls (1.1,-0.6) and (0.5,-0.8) .. (0.3,-0.1);
\draw[thick,shift={(6 cm,0cm)},yscale=-1] (0.2,0.2) .. controls (0,1) .. (0,1) ;
\draw (8.5,0) node {$=\qquad v^{-1}$};
\draw[thick,->,shift={(9.7 cm,0cm)}]  (0,-1.2) .. controls (0.2,0) and (-0.2,0) ..  (0,1) ;
\end{tikzpicture}
\end{center}

Any diagram $\Knot$ is proportional to the trivial knot $\xygraph{!{0;/r0.75pc/:}!{\vcap-}!{\vcap}}$, where the proportionality factor $H(\Knot;t,v)$ is called the framed HOMFLY polynomial. The normalization can be chosen in a such way that 
\begin{equation}
H(\,\xygraph{!{0;/r0.75pc/:}!{\vcap-}!{\vcap}}\,;t,v)=\frac{v-v^{-1}}{t-t^{-1}}.
\end{equation}
We stress that, according to our definition, $H(\Knot;t,v)$ is an invariant of framed knots; it differs from the original HOMFLY polynomial \cite{Freyd85a} by a multiplicative factor.

We denote by $\mathscr P$ the set of all partitions. Here a partition is a non-increasing sequence of nonnegative integers. The natural basis $\{Q_\lambda\}_{\lambda\in\mathscr P}$ of the HOMFLY skein identified by Morton \cite{Morton02a} furnishes a definition for the colored HOMFLY polynomial
$H_\lambda(\Knot)=H(\Knot * Q_\lambda),$
where $\Knot*Q_\lambda$ is the satellite knot obtained by embedding $Q_\lambda$ into a tubular neighborhood of $\Knot$. Alternatively, $H_\lambda(\Knot)$ can be defined from representations of the quantum group $\mathcal U_q(\sl_N)$. Details on the correspondence between the two approaches can be found in \cite{Morton93a}.

\begin{figure}
\centering
\begin{tikzpicture}
\draw (-3.5,0) .. controls (-3.5,2) and (-1.5,2.5) .. (0,2.5);
\draw[xscale=-1] (-3.5,0) .. controls (-3.5,2) and (-1.5,2.5) .. (0,2.5);
\draw[rotate=180] (-3.5,0) .. controls (-3.5,2) and (-1.5,2.5) .. (0,2.5);
\draw[yscale=-1] (-3.5,0) .. controls (-3.5,2) and (-1.5,2.5) .. (0,2.5);
\draw (-2,.2) .. controls (-1.5,-0.3) and (-1,-0.5) .. (0,-.5) .. controls (1,-0.5) and (1.5,-0.3) .. (2,0.2);
\draw (-1.75,0) .. controls (-1.5,0.3) and (-1,0.5) .. (0,.5) .. controls (1,0.5) and (1.5,0.3) .. (1.75,0);
\draw[thick,color=blue,dashed] (0,-2.5) .. controls (-0.75,-2.5) and (-0.75,-0.5) .. (0,-0.5);
\draw[thick,color=blue,xscale=-1] (0,-2.5) .. controls (-0.75,-2.5) and (-0.75,-0.5) .. (0,-0.5);
\draw[thick,color=blue,->] (0.55,-1.3) .. controls (0.55,-1.3) .. (0.55,-1.3);
\draw (-1,-1.25) node [left] {$\beta$};
\draw (0.6,-1.1) node [right] {$\alpha$};
\draw[thick,color=red,dashed] (-3.5,0) .. controls (-3.5,1) and (-1.5,1.75) .. (0,1.75);
\draw[thick,color=red,dashed,xscale=-1] (-3.5,0) .. controls (-3.5,1) and (-1.5,1.75) .. (0,1.75);
\draw[thick,color=red,rotate=180] (-3.5,0) .. controls (-3.5,1) and (-1.5,1.75) .. (0,1.75);
\draw[thick,color=red,yscale=-1,->] (-3.5,0) .. controls (-3.5,1) and (-1.5,1.75) .. (0,1.75);
\end{tikzpicture}
\caption{Basis of the first homology group $H_1(\Torus^2)$ of the torus.}\label{homology_t2}
\end{figure}
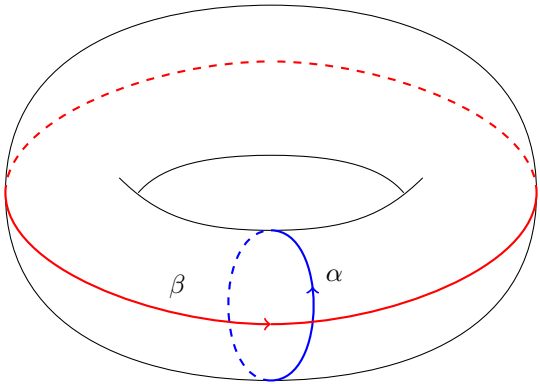

Let $\Torus^2$ denote the torus and fix a basis $\{\alpha,\beta\}$ of the first homology group $H_1(\Torus^2)$ (see Fig. \ref{homology_t2}). Any simple oriented closed curve $\gamma$ on $\Torus^2$ is completely determined, up to isotopy, by two relatively prime integers $n,m\in\ZZ$ such that
\begin{equation}
[\gamma]=m[\alpha]+n[\beta].
\end{equation}
The isotopy class of those curves determined by $(n,m)$ is called the $(n,m)$-torus knot and denoted $\Torus(n,m)$. 
In general, no closed formula is known for $H_\lambda(\Knot)$. For torus knots however, an explicit formula has been derived by Lin \& Zheng \cite{Lin06a} by using Hecke algebras. Their formula involve the Adams coefficients $c_{\lambda,n}^{\nu}\in\ZZ$ defined by
\begin{equation}
s_\lambda(\vec x^n)=\sum_{\nu\in\mathscr P} c_{\lambda,n}^\nu s_\nu(\vec x),
\end{equation}
where $s_\lambda(\vec x)$ is the Schur polynomial. The result, which can be seen as a generalization of the Rosso-Jones formula \cite{Rosso93a}, reads
\begin{equation}
H_\lambda\big(\Torus(n,m);t,v\big)=v^{n|\lambda|}\sum_{\nu\in\mathscr P}c_{\lambda,n}^\nu t^{\frac m n \varkappa_\nu} d_\nu(t,v)\,,\label{rosso-jones}
\end{equation}
where $d_\nu(t,v)= H(Q_\nu;t,v)$ are the invariants of the unknot, which are usually referred to as quantum dimensions. The combinatorial quantity
\begin{equation}
\varkappa_\lambda=\sum_{i=1}^{\ell(\lambda)}\sum_{j=1}^{\lambda_i}(j-i)
\end{equation}
is related to the quadratic Casimir of $SU(N)$.
\subsection{Knots in Lens Spaces}\label{lens-space}
For $(p,q)\in\ZZ^*_+\times\ZZ$, the lens space $L(p,q)$ is the quotient of $\S^3$ (considered as the unit sphere in $\C^2$) by the $\ZZ_p$-action
\begin{equation}
\begin{array}{rcl}
\ZZ_p \times \C^2 & \longrightarrow & \C^2\\
\big([1]_p,(z,w)\big)&\longmapsto& \big(\e^{2\pi i/p}z,\e^{2\pi iq/p}w\big).
\end{array}
\end{equation}
In the polar representation $(z,w)=(r \e^{i\theta},\sqrt{1-r^2}\e^{i\phi})$ of points in $\S^3$, the fundamental domain is
\begin{equation}
\Big\{(r,\theta,\phi):r\in[0,1],\quad\theta\in[0,2\pi[,\quad\phi\in[0,\frac{2\pi}{p}[\Big\}.
\end{equation}
Geometrically, $L(p,q)$ is obtained by gluing two solid tori $\mathcal T_1,\mathcal T_2\approx\D^2\times\S^1$ along their boundary 
\begin{equation}
 L(p,q) = \mathcal T_2 \cup_{\phi_{p,q}}\mathcal T_1
\end{equation}
using the homeomorphism
\begin{equation}
\phi_{p,q}=\begin{pmatrix}
p & r\\
q & s
\end{pmatrix}\in PSL(2,\ZZ),
\end{equation}
where $r,s\in\ZZ$ are such that $ps-qr=1$. The common boundary of the two solid tori is called the Heegaard torus; it is unique up to isotopy \cite{Bonahon83a}.
\begin{example}\begin{figure}
\centering
\includegraphics[width=0.5\linewidth]{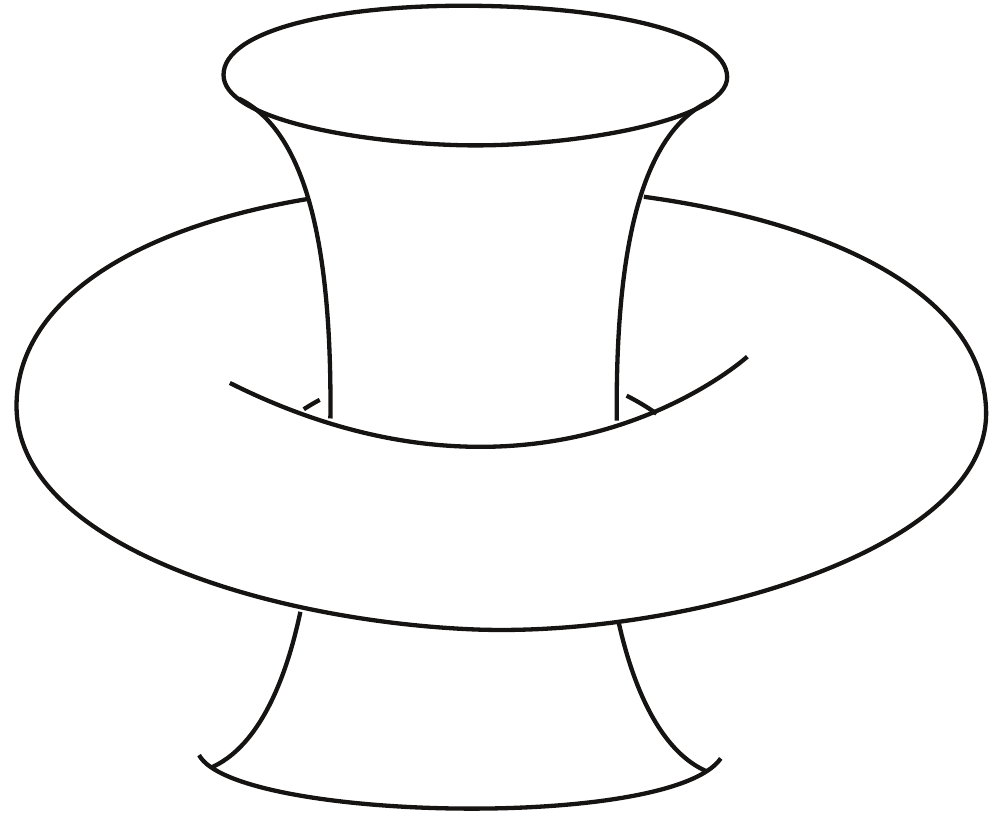}
\caption{Heegaard splitting of $\S^3$ as two solid tori.}\label{heegaard_s3}
\end{figure}\begin{enumerate}
\item The lens space $L(1,q)$ is $\S^3$ for all $q\ge 0$.
Notice that $\phi_{1,0}$ corresponds to the action of
\begin{equation}
\tens S=\begin{pmatrix}0 & -1\\ 1 & 0\end{pmatrix}.
\end{equation}
This Heegaard decomposition of $\S^3$ is represented in Fig. \ref{heegaard_s3}.

\item $\S^2\times\S^1$ is the lens space $L(0,1)$.

\item The $(p,1)$-lens spaces are quotients of $\S^3$ by a cyclic group, $L(p,1)=\S^3/\ZZ_p$. The simplest example is $\R P^3\approx\S^3/\ZZ_2=L(2,1)$.
\end{enumerate}
\end{example}

The mathematical theory of knots in arbitrary $3$-manifolds is not as developed as in $\S^3$. In general, even the existence of an invariant satisfying the skein relation is not clear. For rational homology 3-spheres $M$ which are either atoroidal or Seifert fibered spaces though, Kalfagianni \& Lin \cite{Kalfagianni99a} have showed the existence and uniqueness of an invariant $J(\Knot,M)\in\ZZ[\![t^{\pm 1},v^{\pm 1}]\!]$ satisfying the HOMFLY skein relation.

In general $J(\Knot,M)$ is not a polynomial but a power series. Instead of $(t,v)$, we often use the variables $(\hbar,N)$ defined by $(t,v)=(\e^{\hbar/2},\e^{\hbar N/2})$. In the case of $\R P^3$, an analogue of the Jones polynomial can be constructed \cite{Drobotukhina91a,Prasolov97a}. Recently, Cornwell \cite{Cornwell10a} has proved that for all lens spaces $L(p,q)$ the invariant $J\big(\Knot,L(p,q)\big)$ is a polynomial.

Several comments are in order concerning knots in lens spaces $L(p,q)\neq\S^3$. We will call torus knot a knot which is isotopic to a simple closed curve on the Heegaard torus. Torus knots characterized by different integers $(n,m)$ can be isotopic as knots embedded in $L(p,q)$; an explicit description can be found in \cite{Onaran10a}.

Contrary to $\S^3$, not all knots are null-homotopic in $L(p,q)$ when the fundamental group $\pi_1\big(L(p,q)\big)$ is nontrivial. In particular, there is a trivial knot representing each conjugacy class of $\pi_1\big(L(p,q)\big)$. According to our conventions, the torus knot $\Torus(0,1)$ is the honest unknot (it bounds a disk in $L(p,q)$), whereas $\Torus(1,0)$ is isotopic to the generator $b=[\beta]$ of the fundamental group
\begin{equation}
\pi_1\big(L(p,q)\big)\simeq\av{b|b^p=1}.
\end{equation}
The null-homotopic knots, which lie in an affine part of $L(p,q)$, are called \emph{affine} or \emph{local} knots. They can be identified with knots in $\S^3$. The non null-homotopic unknots, called \emph{rational unknots}, have also been classified \cite{Geiges13a}.

\subsection{Quantum Chern--Simons Theory}\label{tqft}
We now give a brief introduction to the Chern-Simons theory approach to knot invariants. Let $M$ be a smooth, orientable $3$-manifold without boundary. Chern-Simons theory on $M$ is a topological quantum field theory with action given by the Chern-Simon form \cite{Chern74a}
\begin{equation}
S_M(\vec A)=\frac{1}{4\pi}\int_M\Tr\Big[\vec A\wedge\ud\vec A+\frac 2 3\vec A\wedge\vec A\wedge\vec A\Big],
\end{equation}
where $\vec A$ is a $\g$-valued $1$-form on $M$ and $\g$ is the Lie algebra of the gauge group $G=SU(N)$.

The gauge-invariant observables associated with knots $\Knot\subset M$ are Wilson loops
\begin{equation}
W_\lambda^\Knot(\vec A)=\Tr_{V(\lambda)}\Big[\mathcal P\exp\oint_\Knot\vec A\Big],
\end{equation}
where $\mathcal P\exp$ denotes the path-ordered exponential, $\lambda\in\Part$ is a partition, $V(\lambda)$ is the irreducible representation of $SU(N)$ with highest weight represented by $\lambda$, and $\Tr_{V(\lambda)}$ is the trace in the representation $V(\lambda)$. Expectation values of these Wilson loops produce knot invariants:
\begin{equation}
J_\lambda(\Knot,M)=\frac{1}{Z(M)}\int\UD[\vec A]\, W_\lambda^\Knot(\vec A)\e^{ik S_M(\vec A)},
\end{equation}
where $\UD[\vec A]$ is a gauge-invariant measure and $Z(M)$ is the partition function
\begin{equation}
Z(M)=\int\UD[\vec A]\, \e^{ik S_M(\vec A)}.
\end{equation}

For any smooth, orientable, closed surface $\Sigma$, canonical quantization on $\Sigma\times\R$ produces a Hilbert space $\Hilb(\Sigma)$. Witten \cite{Witten89a} showed that $\Hilb(\Sigma)$ is the space of conformal blocks of the Wess-Zumino-Witten model on $\Sigma$. When $\Sigma$ is the torus $\Torus^2$, the space $\Hilb(\Torus^2)$ is the set of integrable representations at level $k$ of $G=SU(N)$, but in practice we can choose $k$ sufficiently large and identify $\Hilb(\Torus^2)$ with the representation ring of $SU(N)$. More precisely, 
\begin{equation}
\mathcal H(\Torus^2)=\bigoplus_{\lambda\in\mathscr P}\C\ket{\rho+\lambda},
\end{equation}
where $\rho$ is the partition corresponding to the Weyl vector of $SU(N)$. The state $\ket{\rho+\lambda}$ is obtained by placing a Wilson loop in the representation $\lambda$ along the core of a solid torus, and $\ket\rho$ corresponds to the empty state.

This space carries a representation of the modular group $PSL(2,\ZZ)$, so that any self-homeomorphism $\phi$ of $\Torus^2$ can be represented by an operator $\tens U_\phi$ on $\Hilb(\Torus^2)$. If we assume $M$ is a lens space, we can take $\Sigma$ to be the Heegaard torus, and compute the partition function as
\begin{equation}
Z\big(L(p,q)\big)=\bra\rho\tens F_{p,q}\ket\rho,
\end{equation}
where $\tens F_{p,q}=\tens U_{\phi_{p,q}}$. If $\Knot$ is a knot in the interior of a solid torus, we can represent it by a state $\ket{\Knot,\lambda}\in\Hilb(\Torus^2)$, and compute
\begin{equation}
J_\lambda\big(\Knot,L(p,q)\big)=\frac{\bra\rho\tens F_{p,q}\ket{\Knot,\lambda}}{\bra\rho\tens F_{p,q}\ket\rho}.
\end{equation}
For torus knots, Labastida \textit{et al.} \cite{Labastida89a,Labastida91a} explicitly constructed an operator formalism that associates to $\Torus(n,m)$ an operator $\vec W^{\Torus(n,m)}_\lambda$ on $\Hilb(\Torus^2)$. For torus knots in lens spaces, we have the general formula
\begin{equation}
J_\lambda\big(\Knot,L(p,q)\big)=\frac{\bra\rho\tens F_{p,q}\tens W_\lambda^{\Torus(n,m)}\ket{\rho}}{\bra\rho\tens F_{p,q}\ket\rho}.
\end{equation}
Using the explicit representation of torus knot operators \cite{Labastida91a} and group-theoretic consideration, we derived the simple formula \cite{Stevan10a} 
\begin{equation}
\tens W^{\Torus(n,m)}_\lambda\ket{\rho}=\sum_{\mu\in\Part}c_{\lambda,n}^\mu \e^{\frac m n\hbar(N\abs\mu+\varkappa_\mu)}\ket{\rho+\mu},\label{stevan}
\end{equation}
where $c_\lambda^\mu$ and $\varkappa_\mu$ are the same as in \eqref{rosso-jones}.
We thus obtain a compact formula for torus knot invariants.

In the case of $M=\S^3$, the topological invariant $J_\lambda(\Knot,\S^3)$ is up to normalization the colored HOMFLY polynomial $H_\lambda^{\Knot}$. In the case of torus knots, $H_\lambda^{\Torus(n,m)}$ has the explicit expression
\begin{equation}
H_\lambda^{\Torus(n,m)}=\sum_{\mu\in\Part}c_{\lambda,n}^\mu \e^{\frac m n\hbar(N\abs\mu+\varkappa_\mu)}d_\mu,\label{homfly_cs}
\end{equation}
where $d_\mu$ are the invariants of the unknot that appear in \eqref{rosso-jones}. They are obtained in CS theory as
\begin{equation}
d_\lambda=\frac{\bra{\rho}\tens S\ket\lambda}{\bra{\rho}\tens S\ket{\rho}},
\end{equation}
where $\tens S$ here denotes the representative on $\Hilb(\Torus^2)$ of $\tens S\in SL(2,\ZZ)$ given in Example 1$(i)$, and $\bra{\mu}\tens S\ket\lambda$ is given by the Kac--Peterson formula \cite{Kac84a,Gepner86a}. The generalized Rosso--Jones formula \eqref{rosso-jones} then follows from \eqref{stevan}. There are generalizations to torus links \cite{Isidro93a,Stevan10a} and cable knots \cite{Morton08a,Stevan10a}.

According to Witten \cite{Witten89a}, \eqref{homfly_cs} computed at fixed $N$ is a rational function in $t$ by identifying $\e^{\hbar/2}\to t$. Moreover, the family of $1$-variable invariants $H_\lambda^{\Torus(m,n)}|_{\e^{\hbar/2}\to t}$ is equivalent to a $2$-variables invariant through the identification $t^N\to v$ (see \cite{Jones89a} for a discussion).

\section{Chern-Simons Theory on Lens Spaces}\label{cs_lens}
By applying the knot operator formalism along the same lines, we can compute invariants of knots in lens spaces as well. To justify our approach, we briefly discuss the CS partition function of $L(p,1)$, where rigorously proved results are known. In section \ref{unknot_lens} we review the explicit computations for the unknot in $L(p,1)$.

\subsection{Reshetikhin-Turaev Invariant}
The CS partition function of $L(p,1)$ reads
\begin{equation}
Z\big(L(p,1)\big)=\bra{\rho}\tens F_{p,1}\ket{\rho}.
\end{equation}
Up to a normalization factor that will not be needed for our purposes, $\tens F_{p,1}$ is given by \cite{Rozansky96a}
\begin{equation}
\tens F_{p,1}=\sum_{\ell\in\ZZ^N/p\ZZ^N}\tens F^{(\ell)}_{p,1},\label{hansen}
\end{equation}
where
\begin{equation}
\bra\alpha\tens F^{(\ell)}_{p,1}\ket\beta=\sum_{w\in\Sym_N}(-1)^w\exp\Big[\frac{\hbar}{p}\Big(\alpha^2-2\alpha(k'\ell+w(\beta)\big)+(k'\ell+w(\beta)\big)^2\Big)\Big].
\end{equation}
The explicit expression for $Z\big(L(p,1)\big)$ obtained from this formula coincides with the mathematical result \cite{Hansen04a} for the Reshetikhin-Turaev invariant of $L(p,1)$.

In the mathematical literature, \eqref{hansen} is usually computed for fixed $N$ and considered a polynomial in $t'=\e^{\hbar'}$, where $\hbar'=\hbar/p$. If one considers $N$ as a formal parameter, which will be natural in order to generalize the colored HOMFLY polynomial, there is no consistent way to identify $Z\big(L(p,1)\big)$ as a two-variable polynomial or power series.

By studying $Z\big(L(p,1)\big)$ in the context of topological string theory, Aga\-na\-gic \textit{et al.} \cite{Aganagic04a} considered the summands
\begin{equation}
\bra\alpha\tens F^{(\ell)}_{p,1}\ket\beta= \sum_{w\in\Sym_N}(-1)^w\exp\Big[\hbar'\Big(\alpha^2-2\alpha(k'\ell+w(\beta)\big)+(k'\ell+w(\beta)\big)^2\Big)\Big]
\end{equation}
separately, and in particular the partition function $Z_\ell\big(L(p,1)\big)=\bra0\tens F^{(\ell)}_{p,1}\ket0$. The quotient
\begin{equation}
\ZZ^N/(p\ZZ^N \ltimes \Sym_N)=\bigsqcup_{N_1+\dots+N_p=N} C_{N_1,\dots,N_p}
\end{equation}
has a decomposition in classes
\begin{equation}
C_{N_1,\dots,N_p}=\Sym_N\cdot\Big((\underbrace{0,0,\dots,0}_{N_1},\underbrace{1,1,\dots,1}_{N_{2}},\dots,\underbrace{p-1,\dots,p-1}_{N_p})+p\ZZ^N\Big).
\end{equation}
It follows that the contribution $\bra\alpha\tens F^{(\ell)}_{p,1}\ket\beta$ corresponding to $\ell\in C_{N_1,\dots,N_p}$ has a well-defined asymptotic expansion in the $p+1$ variables $\hbar'$ and $N_1,\dots,N_p$. Moreover $Z_\ell\big(L(p,1)\big)$ can be written as the matrix integral \cite{Aganagic04a}
\begin{equation}
Z_\ell\big(L(p,1)\big)=\int\ud^N x\,\e^{-\frac{1}{2\hbar'}\vec x^2-\ell\cdot \vec x}\prod_{i<j}\Big(2\sinh\frac{x_i-x_j}{2}\Big)^2\label{Zlens}
\end{equation}
up to normalization. Matrix models with this deformed version of the Vandermonde determinant first appeared in \cite{Marino05c} to compute partition functions of more general Seifert manifolds. This expression has also been derived by means of localization techniques \cite{Beasley05a}. The behavior of the full partition function $Z\big(L(p,1)\big)$ can also be studied from the point of view of matrix models and involves instanton contributions \cite{Marino10b}.

For $q\neq 1$, the partition function can also be written as a matrix integral \cite{Brini08a}
\begin{multline}
Z_\ell\big(L(p,q)\big)=\int\ud^N x\,\e^{-\frac{1}{2\hbar'}\vec x^2-\ell\cdot\vec x}\prod_{i<j}2\sinh\Big(\frac{x_i-x_j}{2}\Big)\\
\times\prod_{i<j}2\sinh\Big(\frac{x_i-x_j}{2}+\frac{q-1}{p}(\ell_i-\ell_j)\Big),
\end{multline}
which reduces to \eqref{Zlens} when $q=1$. From the point of view of CS theory, the relationship with topological string theory and therefore the algebraic curve description is not expected to hold in this case \cite{Auckly06a}. Hence we restrict to $L(p,1)$ in the sequel, and perform explicit computations for $L(2,1)$ only.

\subsection{Unknot}\label{unknot_lens}
As a first step towards our analysis of torus knots in $L(p,1)$, we recall the results of Bouchard \textit{et al.} \cite{Bouchard09a} on the unknot. Along the same lines of reasoning as \cite{Aganagic04a}, they consider a family of quantities
\begin{equation}
J_{\ell,\lambda}\big(\Torus(1,0),L(p,1);\hbar,N\big)= \frac{\bra{\rho}\tens F^{(\ell)}_{p,1}\ket{\rho+\lambda}}{\bra{\rho}\tens F^{(\ell)}_{p,1}\ket{\rho}}\label{unknot_rp3}
\end{equation}
indexed by $\ell\in\ZZ^N/p\ZZ^N$, such that
\begin{equation}
J_{\lambda}\big(\Torus(1,0),L(p,1)\big)=\sum_{\ell\in\ZZ^N/p\ZZ^N}J_{\ell,\lambda}\big(\Torus(1,0),L(p,1)\big).
\end{equation}
Though not specified, their computations correspond to the rational unknot $\Torus(1,0)$. Notice that $J_{\ell,\lambda}\big(\Knot,L(p,1)\big)$ is not an invariant of the pair $\big(L(p,1);\Knot)$, but it can be considered an invariant of the triple $\big(L(p,1),\varrho;\Knot\big)$, where $\varrho:\pi_1\big(L(p,1)\big)\longrightarrow SU(N)$ is the homomorphism (or equivalently the choice of a flat connection) determined by $\ell$.

These quantities \eqref{unknot_rp3} generalize the quantum dimension in the sense that
\begin{equation}
J_{0,\lambda}\big(\Torus(1,0),L(p,1);\hbar,N\big)=\e^{\hbar' (N\abs\lambda+\varkappa_\lambda)}d_\lambda(\hbar',N)=H_{\lambda}^{\Torus(1,1)}(\hbar',N).\label{rp3-qdim}
\end{equation}
This result follows directly from the observation that
\begin{align}
 \bra\alpha\tens F^{(0)}_{p,1}\ket\beta&=\sum_{w\in\W}(-1)^w\exp\Big[\hbar'\big(\alpha^2-2\alpha\cdot w(\beta)+ \beta^2\big)\Big]\nonumber\\
 &=\bra\alpha\tens F_{(1,1)}\ket\beta\Big|_{\hbar\to\hbar'}.
\end{align}
As for the partition function, the unknot invariants can be expressed as a matrix integral \cite{Bouchard09a}
\begin{equation}
\bra{\rho}\tens F^{(\ell)}_{p,1}\ket{\rho+\lambda}=\int\ud^N x\,\e^{-\frac{1}{\hbar'} \vec x^2-\ell\cdot\vec x}\prod_{i<j}\Big(2\sinh\frac{x_i-x_j}{2}\Big)^2 s_\lambda(\e^{\vec x}).\label{mm-unknot}
\end{equation}
This formula has also been derived using localization techniques \cite{Gang09a}.

Integrals of the form \eqref{mm-unknot} can be rewritten as an effective multi-matrix model and computed perturbatively \cite{Aganagic04a,Bouchard09a}. In principle it is possible to compute $J_{\ell,\lambda}\big(\Torus(1,0);L(p,1)\big)$ for arbitrary $p$ as an asymptotic series in $p+1$ variables. For definiteness, we perform actual computations only for $L(2,1)=\R P^3$. In the sequel, we denote by $R_\lambda^{\Knot}(N_1,N_2)$ the series $J_{\ell,\lambda}\big(\Knot;\R P^3\big)$ corresponding to
\begin{equation}
\ell=(\underbrace{1,1,\dots,1}_{N_1},\underbrace{0,0,\dots,0}_{N_2}).
\end{equation}
To compute this invariant at finite order in $\hbar'$, we proceed as follows. On the one hand, $R_\lambda^{\Knot}(N_1,N_2)$ is a polynomial with rational coefficients $R_\lambda^{\Knot}(N_1,N_2)\in\Q[\hbar',N_1,N_2]$. On the other hand, by expanding \eqref{mm-unknot} in series we obtain a sum of Gaussian integrals, which can be evaluated numerically at fixed values of $N_1$ and $N_2$. Thus $R_\lambda^{\Knot}(N_1,N_2)$ is retrieved as an interpolation polynomial by performing sufficiently many numerical evaluations.
\begin{align}
R^{\Torus(1,0)}_{\Yboxdim4pt\yng(1)}(N_1,N_2)&=(N_1-N_2)+\frac{1}{2} \left(N_1^2-N_2^2\right)\hbar'\nonumber\\
&+\frac{1}{24} \left(N_2-N_1+4 N_1^3+6 N_1^2 N_2-6 N_1 N_2^2-4 N_2^3\right)\hbar'^2\\
& +\frac{1}{48} \left(N_2^2-N_1^2+2 N_1^4+6 N_1^3 N_2-6 N_1 N_2^3-2 N_2^4\right)\hbar'^3\nonumber\\
&+ \O(\hbar'^4)\nonumber\\
R^{\Torus(1,0)}_{\Yboxdim4pt\yng(2)}(N_1,N_2)&=\Big[\frac{1}{2} (N_1+N_2) +\frac{1}{2} (N_1-N_2)^2 \Big]\nonumber\\
&+\Big[\frac{1}{2} (N_1+N_2) +\left(N_1^2+N_2^2\right)+ \frac{1}{2} (N_1-N_2)^2 (N_1+N_2) \Big] \hbar'\nonumber\\
&+\Big[\frac{1}{6} (N_1+N_2) +\frac{1}{24} \left(17 N_1^2+26 N_1 N_2+17 N_2^2\right)\\
&\qquad+\frac{1}{6} \left(5 N_1^3+3 N_1^2 N_2+3 N_1 N_2^2+5 N_2^3\right)\nonumber\\
&\qquad +\frac{1}{24} (N_1-N_2)^2 \left(7 N_1^2+16 N_1 N_2+7 N_2^2\right)\Big] \hbar'^2+\O(\hbar'^3)\nonumber
\end{align}
These results coincide with those obtained by Bouchard \textit{et al.} \cite{Bouchard09a}. Notice that $R_\lambda^{\Torus(1,0)}(N_1,0)$, which corresponds to $\ell=0$, coincides with $H_\lambda^{\Torus(1,1)}(\hbar',N_1)$. 

\subsection{Torus Knots}
We generalize the results of Bouchard \textit{et al.} \cite{Bouchard09a} to torus knots by extending the generalized Rosso-Jones formula to lens spaces. We consider for the $(n,m)$-torus knot in $L(p,q)$ the quantity
\begin{equation}
 J_{\ell,\lambda}\big(\Torus(n,m);L(p,q)\big)=\frac{\bra{\rho}\tens F_{p,q}^{(\ell)}\tens W^{(n,m)}_\lambda\ket{\rho+\lambda}}{\bra{\rho}\tens F_{p,q}^{(\ell)}\ket{\rho}},
\end{equation}
which depends on $\ell\in\ZZ^N/p\ZZ^N$. Using formula \eqref{stevan} to decompose the knot operator, we obtain
\begin{equation}
J_{\ell,\lambda}\big(\Torus(n,m);L(p,q)\big)=\sum_{\nu\in\Part} c_{\lambda,n}^\nu \e^{\frac m n \hbar (N\abs\nu+\varkappa_\nu)}J_{\ell,\nu}\big(\Torus(1,0);L(p,q)\big)\label{rosso-lens}
\end{equation}
for any lens space $L(p,q)$. Using the matrix integral \eqref{mm-unknot}, this gives a tractable expression for torus knots in $L(p,1)$.

In view of \eqref{rp3-qdim}, one can ask how $J_{\ell,\lambda}\big(\Knot,L(p,1)\big)$ relates to the colored HOMFLY polynomial. Indeed,
\begin{align}
J_{0,\lambda}\big(\Torus(n,m),L(p,1);\hbar,N\big)&=\sum c_{\lambda,n}^\nu \e^{\hbar\frac m n(N\abs\nu+\varkappa_\nu)}\e^{\hbar' (N\abs\nu+\varkappa_\nu)}d_\nu(\hbar',N)\nonumber \\
&=\sum c_{\lambda,n}^\nu \e^{\hbar'\frac{pm+n}{n}(N\abs\nu+\varkappa_\nu)}d_\nu(\hbar',N)\label{relS3RP3}\\
&=H^{\Torus(n,pm+n)}_\lambda(\hbar',N),\nonumber
\end{align}
which reduces to \eqref{rp3-qdim} for $(n,m)=(1,0)$. There is a correspondence between the two invariants in the sense that the former reduces to the latter in a certain limit, however the invariants do not correspond to the same knot. If $\Knot\subset L(p,1)$ is a local knot, it can be considered a knot in $\S^3$ and we expect the two invariants to coincide in the limit $\ell=0$. Schematically, we have a commutative diagram.
\begin{equation}
\xymatrix{\Torus(1,0)\subset L(p,1)\ar[rrr]^{\ell\to 0} \ar[dd]&&&\Torus(1,1)\subset\S^3\ar[dd] \\
\\
 \Torus(n,m)\subset L(p,1)\ar[rrr]^{\ell\to 0} &&& \Torus(n,pn+m)\subset\S^3}
 \end{equation}

Using the same convention as above, we denote $J_{\ell,\lambda}\big(\Torus(n,m),\R P^3\big)$ by $R_\lambda^{\Torus(n,m)}(N_1,N_2)$ for knots in $\R P^3$. Equation \eqref{rosso-lens} furnishes explicit results for $\Torus(2,m)$ and $\Torus(3,m)$.
\begin{align}
R^{\Torus(2,m)}_{\Yboxdim4pt\yng(1)}(N_1,N_2)&=(N_1+N_2)+2 (1+m) \left(N_1^2+N_2^2\right) \hbar'\nonumber\\
&+\frac{1}{6}  (2 + 6 m + 3 m^2)  (N_1+N_2) \hbar'^2\nonumber\\
&+\frac{1}{12} (1 + m)\big[(5 + 16 m + 8 m^2) N_1^2\\
&\qquad+2(9 + 8 m + 4 m^2) N_1 N_2+(5 + 16 m + 8 m^2) N_2^2\big]\hbar'^3\nonumber\\
&+\frac{1}{360}(-8 + 60 m^2 + 60 m^3 + 15 m^4) (N_1 + N_2)\hbar'^4+\O(\hbar'^5)\nonumber\\
R^{\Torus(2,m)}_{\Yboxdim4pt\yng(2)}(N_1,N_2)&=\frac{1}{2}\big((N_1 + N_2) + (N_1^2 + 2N_1 N_2 + N_2^2)\big)\\
&+2(m+1)\big((N_1+N_2)+2(N_1^2+N_2^2)\big)\hbar'+\O(\hbar'^2)\nonumber\\
R^{\Torus(3,m)}_{\Yboxdim4pt\yng(1)}(N_1,N_2)&=(N_1-N_2) +\frac{3}{2} (3+2 m) \left(N_1^2-N_2^2\right)\hbar'\nonumber\\
&+\frac{1}{24} (63 + 96 m + 32 m^2) (N_1 - N_2)\hbar'^2\\
&+\frac{1}{48} (375 + 826 m + 576 m^2 + 128 m^3) (N_1^2 - N_2^2)+\O(\hbar'^3)\nonumber
\end{align}
We can explicitly check that $R_\lambda^{\Torus(n,m)}$ reduces to $H_\lambda^{\Torus(n,2m+n)}$ for $N_2=0$.

\section{Algebraic Curves for Torus Knots in $\S^3$}\label{b-modelS3}
In this section, we review the topological string theory description of torus knots in $\S^3$. With a view to the case of $\R P^3$, we reproduce the results of \cite{Brini11a}.

The intimate relationship between CS theory and topological string theory produces nontrivial results in knot theory. Arising from the large $N$ expansion of 't Hooft  \cite{tHooft74a}, the Gopakumar-Vafa duality conjecture \cite{Gopakumar99a,Auckly06a} relates topological string theory on the resolved conifold
\begin{equation}
\O(-1)\oplus\O(-1)\to\mathbb P^1
\end{equation}
to CS theory on $\S^3$. Formulated for the partition function (the Reshetikhin--Turaev invariant) at first \cite{Gopakumar99a}, the conjecture has been generalized to knots by incorporating Lagrangian submanifolds \cite{Ooguri00a}. In this enumerative description (A-model), the conjecture provides reformulations in terms of integer invariants for the knot polynomials. There are proposals for the Lagrangian submanifold that describes torus knots \cite{Labastida00a,Taubes06a,Diaconescu11a,Jockers12a} and indirect evidences based on integrality properties \cite{Labastida01a}.

By mirror symmetry \cite{Hori03a}, there exists a different description (B-model) in terms of non-compact Riemann surfaces embedded in $\C^*\times \C^*$. This description is known for the unframed unknot \cite{Aganagic00a} and the framed unknot \cite{Aganagic02a}. As proposed first by Mari\~no \cite{Marino08b} and developed by Bouchard \textit{et al.} \cite{Bouchard09a}, the computation of amplitudes involves a recursive procedure adapted from the formalism of Eynard \& Orantin \cite{Eynard07a}. Recently, Brini, Eynard \& Mari\~no \cite{Brini11a} have found such a description for torus knots in $\S^3$. More generally, this universal procedure should allow to compute all colored HOMFLY invariants of any knot $\Knot$ embedded in $\S^3$ from a non-compact algebraic curve. According to Aganagic \& Vafa \cite{Aganagic12a}, the algebraic curve is related to the $A$-polynomial of $\Knot$.

Before detailing the recursive procedure, we introduce generating functions and reformulated invariants that make the identification manifest.  
Let
\begin{equation}
\bar\Xi(\vec z)=\sum_{\lambda\in\Part} H_{\lambda}^{\Knot}(t,v)s_\lambda(\vec z)
\end{equation}
be the generating function for invariants of the $(n,m)$-torus knot in $\S^3$. The cumulants $\hat H^\Knot_{\vec k,g}(v)$ are polynomial in $v^{\pm 1}$ defined by
\begin{equation}
\log\bar\Xi=\sum_{\vec k\in\N^\N}\frac{1}{\mathfrak z_{\vec k}} \hat H^{\Knot}_{\vec k} (t,v)p_{\vec k}(\vec z),
\end{equation}
where $p_{\vec k}$ are the Newton polynomials and $\mathfrak z_{\vec k}=\prod_j k_j!j^{k_j}$. The cumulants (referred to as connected expectation values in the physics literature) posses an asymptotic expansion
\begin{equation}
\hat H_{\vec k}^{\Knot}(t,v)=\sum_{g\ge 0}\hbar^{2g-2+\abs{\vec k}} \hat H^{\Knot}_{\vec k,g} (v).
\end{equation}
To summarize, $\log\bar\Xi$ has the structure
\begin{equation}
\log \bar\Xi=\sum_{g\ge 0}\sum_{\vec k\in\N^\N}\frac{1}{\mathfrak z_{\vec k}}\hbar^{2g-2+\abs{\vec k}} \hat H^{\Knot}_{\vec k,g} (v)p_{\vec k}(\vec z).
\end{equation}
The first cumulants can be computed by hand, for instance
\begin{gather*}
\hat H_{(1)}^{\Knot}=H^{\Knot}_{\Yboxdim4pt\yng(1)}\qquad\qquad\hat H_{(0,1)}^{\Knot}=H^{\Knot}_{\Yboxdim4pt\yng(2)} -H^{\Knot}_{\Yboxdim4pt\yng(1,1)}\\
\hat H_{(2)}^{\Knot}=H^{\Knot}_{\Yboxdim4pt\yng(2)}+H^{\Knot}_{\Yboxdim4pt\yng(1,1)}-(H^{\Knot}_{\Yboxdim4pt\yng(1)})^2.
\end{gather*}

\subsection{Topological Recursion}
For our purposes, a spectral curve is a smooth complex algebraic curve
\begin{equation}
\Sigma=\big\{\mathcal E(x,y)=0\big\}\subset\C^*\times\C^*.
\end{equation}
Equivalently, it can be defined by a non-compact Riemann surface $\Sigma$ with meromorphic coordinates $x,y:\Sigma\longrightarrow\C^*$. We assume that $x:\Sigma\longrightarrow\C$ has only simple ramification points. Different choices of coordinates on $\Sigma$ correspond to different embeddings of the Riemann surface, which are encoded in the polynomial $\mathcal E(x,y)$.

The topological recursion defines a family of symmetric meromorphic differentials
\begin{equation}
\Omega_{g,n}(p_1,\dots,p_n)\in\mathrm{Sym}^n\big(\Omega^1(\Sigma)\big)\qquad \text{for }g\ge 0 \text{ and } n\ge 1
\end{equation}
by a recursive procedure. Before we formulate the recursion, we need some ingredients.  Given a spectral curve in a particular parameterization, we define a $1$-form
\begin{equation}
\Theta(p)=\log y(p)\frac{\ud x(p)}{x(p)}.
\end{equation}
We also define the Bergman kernel as follows: $B(z_1,z_2)$ is the symmetric meromorphic differential form on $\Sigma\times\Sigma$  with a double pole along the diagonal $z_1=z_2$, no residue, and no other pole. It is uniquely determined by imposing a normalization condition. We finally need the $1$-form
\begin{equation}
\ud E_q(p)=\frac 1 2\int_{z=q}^{z=\bar q} B(p,z).
\end{equation}
Starting from
\begin{equation}
\Omega_{0,1}(p)=\Theta(p)\qquad\text{and}\qquad\Omega_{0,2}(z_1,z_2)=B(z_1,z_2), 
\end{equation}
we define recursively
\begin{multline}
\Omega_{g,h+1}(p,p_1,\dots,p_h)=\sum_{q_i}\Res_{q=q_i}\frac{\ud E_q(p)}{\Theta(p)-\Theta(\bar q)}\Bigg[ \Omega_{g-1,h+2}(q,\bar q,p_1,\dots,p_h)\\
+\sum_{l=0}^g\sum_{\emptyset\neq J\subsetneq H} \Omega_{g-l,\abs J}(q,p_J)\Omega_{l,h-\abs J}(\bar q,p_{H\setminus J})\Bigg],\label{top-rec}
\end{multline}
where $\{q_i\}$ is the set of ramification points of $\mathcal E$.

In the sequel, we will denote by $\bar{\mathcal E}_\Knot$ the curve corresponding to $\Knot$ in $\S^3$ and by $\bar\Omega_{g,h}^{\Knot}$ the meromorphic differentials obtained by applying the topological recursion to $\bar{\mathcal E}_\Knot$. Let us finally define
\begin{equation}
\bar\omega^\Knot_{g,h}(z_1,\dots,z_h)=\bar\Omega^\Knot_{g,h}(z_1,\dots,z_h)-\delta_{g,0}\delta_{h,2}\frac{\ud z_1\ud z_2}{(z_1-z_2)^2}.
\end{equation}
The conjecture due to Brini, Mari\~no \& Eynard \cite{Brini11a} reads
\begin{equation}
\int\bar\omega_{g,h}^{\Torus(n,m)}=\sum_{\abs{\vec k}=h}\frac{1}{\mathfrak z_{\vec k}}\hat H^{\Torus(n,m)}_{\vec k,g} (v)p_{\vec k}(\vec z).\label{conjS3}
\end{equation}

\subsection{Unknot}
Aganagic \& Vafa \cite{Aganagic00a} determined the algebraic curve
\begin{equation}
\bar{\mathcal E}_{\Torus(1,0)}(x,y)=y-xy+v^2x-1\label{E_unknot1}
\end{equation}
describing the (unframed) unknot in $\S^3$. Through the remodeled B-model formalism of Bouchard \textit{et al.} \cite{Bouchard09a}, all HOMFLY invariants can be obtained by applying the topological recursion \eqref{top-rec} to this curve. More generally, the framed unknot $\Torus(1,f)$ is described by the curve \cite{Aganagic02a}
\begin{equation}
\bar{\mathcal E}_{\Torus(1,f)}(x,y)=y^{f+1}-y^f-xy+v^2x .
\end{equation}
Although one usually considers $\Torus(1,0)$ as the standard trivial knot in $\S^3$ and correspondingly use \eqref{E_unknot1} as the spectral curve, to facilitate the comparison between $\S^3$ and $\R P^3$ it is more convenient to start from
\begin{equation}
\bar{\mathcal E}_{\Torus(1,1)}(x,y)=y^2-y(1+x)+v^2x.\label{E_unknot}
\end{equation}
In view of \eqref{rp3-qdim}, we already know that $\Torus(1,0)$ in $\R P^3$ corresponds to $\Torus(1,1)$ in $\S^3$.

Locally, \eqref{E_unknot} defines $y$ as a function of $x$,
\begin{equation}
y_{\Torus(1,1)}(x)=\frac{(1+x)+\sqrt{(1+x)^2-4v^2 x}}{2},
\end{equation}
which yields the disk amplitude $\bar\omega_{0,1}^{\Torus(1,1)}(x)=\log y_{\Torus(1,1)}(x)\frac{\ud x}{x}$. After integration
\begin{align}
\int\bar\omega_{0,1}^{\Torus(1,1)}=(v^2-1)x+\frac{3v^4-4v^2+1}{4}x^2+\frac{10v^6-18v^4+9v^2-1}{9}x^3+\dots
\end{align}
we precisely reproduce the unknot invariants
\begin{align}
\hat H_{(1);0}^{\Torus(1,1)}&=v^2-1=S  + \frac{S^2}{2} + \frac{S^3}{6} + \frac{S^4}{24} +\O(S^5)\nonumber\\
\hat H_{(0,1);0}^{\Torus(1,1)}&=\frac{3v^4-4v^2+1}{2}=S  + 2 S^2 + \frac{5 S^3}{3} + \frac{11 S^4}{12}+\O(S^5)\\
\hat H_{(0,0,1);0}^{\Torus(1,1)}&=\frac{10v^6-18v^4+9v^2-1}{3}=S + \frac{9 S^2}{2} + \frac{15 S^3}{2} + \frac{59 S^4}{8} +\O(S^5)\nonumber
\end{align}
computed from Chern-Simons theory. Here we have expanded $v=\e^{S/2}$ in power series to help comparing these values with $\R P^3$ invariants.

For knots in $\S^3$, the annulus amplitude $\bar\omega_{0,2}^{\Torus(1,1)}(x_1,x_2)$ is simply given by
\begin{equation}
\bar\omega_{0,2}^{\Torus(1,1)}(x_1,x_2)=\frac{\ud y(x_1)\ud y(x_2)}{\big(y(x_1)-y(x_2)\big)^2}-\frac{\ud x_1\ud x_2}{(x_1-x_2)^2},
\end{equation}
because the spectral curve is topologically the Riemann sphere, hence the Bergman kernel is
\begin{equation}
B(p_1,p_2)=\frac{\ud y(p_1)\ud y(p_2)}{\big(y(p_1)-y(p_2)\big)^2}.\label{bergman}
\end{equation}
One can check that
\begin{equation}
\int\bar\omega_{0,2}^{\Torus(1,1)}(x_1,x_2)=\big[v^2(v^2-1)\big]x_1x_2+\big[2 v^6 - 3 v^4 + v^2\big](x_1^2x_2+x_1x_2^2)+\O(|\vec x|^4)
\end{equation}
reproduces the unknot invariants
\begin{align}
\hat H_{(2),0}^{\Torus(1,1)}&=v^2(v^2-1)=S + \frac{3 S^2}{2} + \frac{7 S^3}{6} + \frac{5 S^4}{8}+\O(S^5)\\
\hat H_{(2,1),0}^{\Torus(1,1)}&=2 v^6 - 3 v^4 + v^2=S + \frac{7 S^2}{2} + \frac{31 S^3}{6} + \frac{115 S^4}{24}+\O(S^5)
\end{align}
Higher order amplitudes can be computed by means of the topological recursion \eqref{top-rec}.

\subsection{Framed Unknot}
According to the proposal of Brini \textit{et al.} \cite{Brini11a}, the $SL(2,\ZZ)$-action on torus knots should correspond to the $SL(2,\ZZ)$ re-parameterization of the spectral curve. But before we turn to the general case, it is instructive to consider an $SL(2,\ZZ)$-action of the form
\begin{equation}
\begin{pmatrix}1 & f\\0 & 1\end{pmatrix}\in SL(2,\ZZ),\label{sl2_framing}
\end{equation}
and the corresponding change of variables
\begin{equation}
(x,y)\longrightarrow (z,y)=(xy^f,y).
\end{equation}
This transformation describes the framed unknot
\begin{equation}
\bar{\mathcal E}_{\Torus(1,f)}(z,y)=y^{f+1}-y^f-zy+v^2z ,
\end{equation}
because the re-parameterized spectral curve
\begin{equation}
\bar{\mathcal E}_{\Torus(1,0)}(zy^{-f},y)=y-zy^{1-f}+v^2z y^{-f}-1
\end{equation}
coincides with $\bar{\mathcal E}_{\Torus(1,f)}$.

The framed amplitudes $\bar\Omega_{g,h}^{\Torus(1,f)}$ can be computed through the topological recursion \eqref{top-rec} by solving $\bar{\mathcal E}_{\Torus(1,f)}$ for $y_{\Torus(1,f)}(x)$. The topological recursion is more easily applied when $\bar{\mathcal E}_{\Knot}$ is hyperelliptic. Therefore, we compute $\bar\omega_{0,1}^{\Torus(1,f+1)}$ by re-parameterizing $\bar\omega_{0,1}^{\Torus(1,1)}$ which is hyperelliptic. Let $x=\bar\eta_f(z)$ be the solution of
\begin{equation}
z=x y(x)^f.
\end{equation}
It is clear that
\begin{equation}
y_{\Torus(1,f+1)}(z)=y_{\Torus(1,1)}\big(\bar\eta_f(z)\big),
\end{equation}
hence
\begin{equation}
\bar\omega_{0,1}^{\Torus(1,f+1)}(z)=\bar\Omega_{0,1}^{\Torus(1,f+1)}(z)=\log y_{\Torus(1,1)}\big(\bar\eta_f(z)\big)\frac{\ud z}{z}.
\end{equation}
Notice that, to check the conjecture, it is enough to know $\bar\eta_f(z)$ as a power series. After integrating $\bar\omega_{0,1}^{\Torus(1,f+1)}(z)$, we obtain
\begin{align}
\int\bar\omega_{0,1}^{\Torus(1,f+1)}&=\big[v^f \left(v^2-1\right)\big] z+\Big[\frac{1}{2} v^{2 f} \left(v^2-1\right) \left(3 v^2+2 f (v^2-1)-1\right)\Big]\frac{z^2}{2}\nonumber\\
&+\Big[\frac{1}{18} v^{3 f} (v^2-1) \big(2-16 v^2+20 v^4\nonumber\\
&\qquad\qquad+9 f^2 (v^2-1)^2+9 f (1-4 v^2+3 v^4)\big)\Big]z^3\\
&+\Big[\frac{1}{48} v^{4 f} (v^2-1) \big(32 f^3(v^2-1)^3+48 f^2 (v^2-1)^2 (3v^2-1)\nonumber\\
&\qquad\qquad+3 (-1+15 v^2-45 v^4+35 v^6)\nonumber\\
&\qquad\qquad+2 f (-11+93 v^2-189 v^4+107 v^6)\big)\Big] z^4+\O(z^5),\nonumber
\end{align}
which reproduces the invariants $\hat H^{\Torus(1,f+1)}_{(0,\dots,0,1);0}$.
\begin{align}
\hat H^{\Torus(1,f+1)}_{(1);0}&=v^f (v^2-1)\nonumber\\
&=S + \frac{1}{2} (1 + f) S^2 + \frac{1}{24} (4 + 6 f + 3 f^2) S^3\\
&\qquad\qquad\qquad + \frac{1}{48} (2 + 4 f + 3 f^2 + f^3) S^4 + \O(S^5)\nonumber\\
\hat H^{\Torus(1,f+1)}_{(0,1);0}&=\frac{1}{2} v^{2 f} \left(v^2-1\right) \left(3 v^2+2 f (v^2-1)-1\right)\nonumber\\
&=S + 2 (1 + f) S^2 + \frac{1}{6} (10 + 18 f + 9 f^2) S^3 \\
&\qquad\qquad\qquad+ \frac{1}{12} (11 + 27 f + 24 f^2 + 8 f^3) S^4 +\O(S^5)\nonumber
\end{align}
The annulus amplitude for the new embedding $\bar{\mathcal E}_{\Torus(1,f+1)}$ is obtained by a simple re-parameterization
\begin{equation}
\bar\Omega_{0,2}^{\Torus(1,f+1)}(z_1,z_2)=\bar\Omega_{0,2}^{\Torus(1,1)}(\eta_f(z_1),\eta_f(z_2)\big),
\end{equation}
because the Bergman kernel does not depend on the particular embedding. After subtraction of the double pole and integration, we obtain
\begin{equation}
\int\bar\omega_{0,2}^{\Torus(1,f+1)}(z_1,z_2)= \Big[\frac 12 (1 + f) v^{2 f} (v^2-1)\big (2 v^2 + f (v^2-1)\big)\Big] z_1z_2+\O(\abs{\vec z}^3)
\end{equation}
which gives the invariant
\begin{align}
\hat H_{(2);0}^{\Torus(1,f+1)}&=\frac 12 (1 + f) v^{2 f} (v^2-1)\big (2 v^2 + f (v^2-1)\big)\nonumber\\
&=(1+f) S+\frac 3 2 (1+f)^2 S^2+\frac{1}{6} \left(7+19 f+18 f^2+6 f^3\right) S^3\\
&\qquad\qquad\qquad\qquad\qquad\quad +\frac{5}{24} (1+f)^2 \left(3+4 f+2 f^2\right) S^4+\O(S^5)\nonumber
\end{align}
Before we move to torus knots, one general comment is in order. Higher amplitudes, that we do not discuss in this work, involve residues at ramification points. Therefore it is not enough to re-parameterize $\bar\Omega_{g,h}^{\Torus(1,1)}$ because the ramification points in the recursion \eqref{top-rec} depend on the embedding  \cite{Bouchard09a}.

\subsection{Torus Knots}
The main result of Brini, Eynard \& Mari\~no \cite{Brini11a} is the proposal of a spectral curve
\begin{equation}
\bar{\mathcal E}_{\Torus(n,m)}(z,y)=y^m(y-1)^n-v^{m-n} z(y-v^2)^n
\end{equation}
such that the topological recursion \eqref{top-rec} reproduces the colored HOMFLY polynomial of the $(n,m)$-torus knot in $\S^3$. This spectral curve has been derived by extending the $SL(2,\ZZ)$-change of variables \eqref{sl2_framing} to $z=x^ny^m$.

Instead of the expected change of variable $z=x^n y(x)^m$, it is more convenient to consider 
\begin{equation}
w=x y(x)^{m/n},
\end{equation}
where $w$ should be thought of as the local coordinate $w=z^{1/n}$ on the curve. Brini \textit{et al.} suggest an additional rescaling $x\to v^{m/n}x$ and define 
\begin{equation}
w=x\Big[\frac{1-v^{m/n+2}x}{1-v^{m/n}x}\Big]^{m/n}.
\end{equation}
As our starting point is $y_{\Torus(1,1)}(x)$ instead of $y_{\Torus(1,0)}(x)$, we slightly adapt their procedure. Let 
\begin{equation}
w=x\Bigg[\frac{(1+v^{m/n}x)+\sqrt{(1+v^{m/n}x)^2-4v^{2+m/n}x}}{2}\Bigg]^{m/n}.
\end{equation}
The inversion $x=\bar\eta_{\frac m n}(w)$ defines a function $\bar\eta_{\frac m n}$ which coincides with $\bar\eta_m$ for $n=1$. The argument of Brini \textit{et al.} implies that $\log y_{\Torus(1,1)}\big(\bar\eta_{\frac m n}(w)\big)$ has a series expansion in integer powers of $w$ (integer powers of $z^{1/n}$) where the integer powers of $w^n$ represent $\bar\omega_{1,0}^{\Torus(n,m+n)}$. In formula, we have
\begin{equation}
\log y_{\Torus(1,1)}\big(\bar\eta_{\frac m n}(w)\big)=\bar\omega_{0,1}^{\Torus(n,m+n)}(w^n)+\sum_{k\neq 0\,\mathrm{mod}\,n}\bar\alpha_{0,1;k}^{\Torus(n,m+n)}w^k.
\end{equation}
In other words, the integer powers of $w^n$ in the curve $\bar{\mathcal E}_{\Torus(1,1)}$ with fractional framing $f=m/n$ can be identified with the integer powers of $z$ in $\bar{\mathcal E}_{\Torus(n,m+n)}$. This description is similar in spirit  to the Rosso-Jones formula \eqref{rosso-jones} for torus knots.

As the analytic behavior of $\bar\eta_{\frac m n}$ depends on $n$, we can only perform computations for fixed $n$. For $n=2$ the invariants of the $(2,m+2)$-torus knot
\begin{align}
\hat H_{(1);0}^{\Torus(2,m+2)}&=\frac{1}{2} v^m (v^2-1) (3 v^2 + m (v^2-1)-1)\nonumber\\
&=S +(2+m) S^2 +\frac{1}{24} \left(40+36 m+9 m^2\right) S^3 \\
&\qquad\qquad\qquad+\frac{1}{24} \left(22+27 m+12 m^2+2 m^3\right) S^4 +\O(S^5)\nonumber\\
\hat H_{(0,1),0}^{\Torus(2,m+2)}&=\frac{1}{12} v^{2 m} (v^2-1) \Big[4 m^3 (v^2-1)^3+12 m^2 (v^2-1)^2 (3 v^2-1)\nonumber\\
&\qquad\qquad\qquad+3 (15 v^2-45 v^4+35 v^6-1)\nonumber\\
&\qquad\qquad\qquad+m (93 v^2-189 v^4+107 v^6-11)\Big]\\
&=S + 4 (2 + m) S^2 + \frac{1}{6} (136 + 132 m + 33 m^2) S^3\nonumber\\
&\qquad\qquad\qquad+ \frac{1}{6} (214 + 299 m + 144 m^2 + 24 m^3) S^4 +\O(S^5)\nonumber
\end{align}
are obtained from
\begin{align}
\int \log y_{\Torus(1,1)}\big(\bar\eta_{\frac m 2}(w)\big)\frac{\ud w}{w}&=\bar\alpha_{0,1;1}^{\Torus(n,m+n)} w+\hat H_{(1);0}^{\Torus(2,m+2)}\frac{w^2}{2}\\
&+\bar\alpha_{0,1;3}^{\Torus(n,m+n)} \frac{w^3}{3}+\hat H_{(0,1);0}^{\Torus(2,m+2)}\frac{w^4}{4}+\O(w^5).\nonumber
\end{align}
One can compute as well for $n=3$ the disk amplitude
\begin{multline}
\int \log y_{\Torus(1,1)}\big(\eta_{\frac m 3}(w)\big)\frac{\ud w}{w}=\bar\alpha_{0,1;1}^{\Torus(3,m+3)} w+\bar\alpha_{0,1;2}^{\Torus(3,m+3)} \frac{w^2}{2}+\hat H_{(1);0}^{\Torus(3,m+3)}\frac{w^3}{3}\\
+\bar\alpha_{0,1;4}^{\Torus(3,m+3)} \frac{w^4}{4}+\bar\alpha_{0,1;5}^{\Torus(3,m+3)} \frac{w^5}{5}+\hat H_{(0,1);0}^{\Torus(3,m+3)}\frac{w^6}{6}+\O(w^7)
\end{multline}
and reproduce the invariants of the $(3,m+3)$-torus knot
\begin{align}
\hat H_{(1);0}^{\Torus(3,m+3)}&=\frac{1}{6} v^m (v^2-1) \big(2-16 v^2+20 v^4\nonumber\\
&\qquad\qquad\qquad\qquad+m^2 (v^2-1)^2+3 m(3 v^4-4 v^2+1)\big)\\
&=S +\frac{3}{2} (3+m) S^2 +\frac{1}{24} \left(180+114 m+19 m^2\right) S^3\nonumber\\
&\qquad\qquad\qquad\qquad +\frac{1}{48} \left(354+316 m+99 m^2+11 m^3\right) S^4 +\O(S^5)\nonumber\\
\hat H_{(0,1);0}^{\Torus(3,m+3)}&=\frac{1}{180} v^{2 m} (v^2-1) \big[8 m^5 (v^2-1)^5+60 m^4 (v^2-1)^4 (3 v^2-1)\nonumber\\
&\qquad +10 m^3(v^2-1)^3 (17-118 v^2+161 v^4)\nonumber\\
&\qquad+15 m^2(v^2-1)^2 (-15+187 v^2-569 v^4+477 v^6)\nonumber\\
&\qquad+30 (-1+35 v^2-280 v^4+840 v^6-1050 v^8+462 v^{10})\\
&\qquad+m (-137+2995 v^2-17660 v^4\nonumber\\
&\qquad\qquad\qquad +41780 v^6-42775 v^8+15797 v^{10})\big]\nonumber\\
&=S + 6 (3 + m) S^2 + \frac{1}{6} (666 + 438 m + 73 m^2) S^3\nonumber\\
&\qquad + \frac{1}{12} (4323 + 4177 m + 1368 m^2 + 152 m^3) S^4 +\O(S^5)\nonumber
\end{align} 
The annulus amplitude is also obtained by simple re-parameterization,
\begin{multline}
\int\Bigg(\Omega_{0,2}^{\Torus(1,1)}\big(\eta_{\frac m 2}(w_1),\eta_{\frac m 2}(w_2)\big)-\frac{\ud w_1\ud w_2}{(w_1-w_2)^2}\Bigg)=\bar\alpha_{0,2;1,1}^{\Torus(2,m+2)}w_1w_2\\
+\bar\alpha_{0,2;2,1}^{\Torus(2,m+2)}\frac{w_1^2w_2+w_2w_1^2}{2}+\hat H_{(2);0}^{\Torus(2,m+2)}\frac{w_1^2w_2^2}{4}+\O(\abs{\vec w}^5).
\end{multline}
\begin{align}
 \hat H_{(2);0}^{\Torus(2,m+2)}&=\frac{1}{4} v^{2 m} (v^2-1) \big[m^4 (v^2-1)^3 + 4 m^3 (v^2-1)^2 (3 v^2-1)\nonumber\\
&\qquad  + 8 v^2 (2 - 9 v^2 + 9 v^4)+ 2 m (-1 + 23 v^2 - 69 v^4 + 51 v^6)\nonumber\\
&\qquad  + m^2 (-5 + 47 v^2 - 95 v^4 + 53 v^6)\big]\\
&= 2 (2 + m) S + 6 (2 + m)^2 S^2\nonumber\\
&\qquad + \frac{1}{3} (176 + 304 m + 150 m^2 + 21 m^3) S^3\nonumber\\
&\qquad  + \frac{1}{12} (1020 + 2628 m + 2227 m^2 + 648 m^3 + 47 m^4) S^4 +\O(S^5)\nonumber
\end{align}

\section{Algebraic Curves for Torus Knots in $\R P^3$}\label{b-modelRP3}
The relationship between CS theory and topological strings also extends to $L(p,1)$ as follows: since $\S^3$ is dual to the resolved conifold, taking a $\ZZ_p$-orbifold on both sides suggests that $L(p,1)$ should be dual to a suitable $\ZZ_p$-quotient of the resolved conifold \cite{Aganagic04a}. The A-model approach to knot invariants is difficult, in particular the determination of the submanifold associated to each knot. Recently, Jockers \textit{et al.} \cite{Jockers12a} have identified the Lagrangian submanifold for torus knots in $\S^3$ and $\R P^3$. It should be interesting to see how their proposal relates with our computations on the CS theory side.

By mirror symmetry \cite{Hori03a}, one can study knots in $L(p,1)$ from the B-model point of view, even without the knowledge of those Lagrangian submanifolds. In particular, Bouchard \textit{et al.} \cite{Bouchard09a} have used their new formalism to test the Gopakumar-Vafa duality for framed unknots. Building on this approach, we propose a B-model description of torus knots in $L(p,1)$ obtained by a $SL(2,\ZZ)$ re-parameterization of the curve for the unknot. As before, we restrict to $L(2,1)\approx\R P^3$. In this case, the A-model geometry is a Calabi-Yau threefold $\mathfrak X$ known as local $\mathbb P^1\times\mathbb P^1$.

The mirror of $\mathfrak X$ is described by a spectral curve, which can be written as
\begin{equation}
\mathcal E_{\Torus(1,0)}(X,Y)=Y^2+(q_sX^2+X+1)Y+q_tX^2,
\end{equation}
where $q_s,q_t$ are parameters. The mirror map identifies these parameters with the 't Hooft parameters, which are (closed) flat coordinates on $\mathfrak X$ near the so-called \emph{orbifold point} \cite{Aganagic04a}:
\begin{equation}
q_t=\frac{1-q_1}{(q_1q_2)^2}\qquad\text{and}\qquad q_s=\frac{1}{(q_1q_2)^2},
\end{equation}
where
\begin{align}
q_1&=2(S_1+S_2)-2(S_1+S_2)^2+\frac{4}{3}(S_1+S_2)^2-\frac 2 3(S_1+S_2)^4+\dots\\
q_2&=\frac{S_1-S_2}{S_1+S_2}+\frac {S_1-S_2}{2}+\frac{S_1^3S_2-S_1S_2^2}{12(S_1+S_2)}+\frac{S_2^3-S_1^3}{24}+\dots
\end{align}

Furthermore, the geometric variables related to the mirror curve have to be expressed in terms of (open) flat coordinates to compare with CS quantities:
\begin{equation}
X=(q_1q_2)x,\label{mirror}
\end{equation}
with $q_1,q_2$ as above. For the sake of clarity, we use the same letter for the geometric variable and the flat coordinate, the former in uppercase and the latter in lowercase.

As for $\S^3$ we denote by $\mathcal E_\Knot$ the curve corresponding to $\Knot$ in $\R P^3$, and introduce the meromorphic differentials $\Omega_{g,h}^{\Knot}$ obtained by applying the topological recursion to $\mathcal E_\Knot$ and
\begin{equation}
\omega^\Knot_{g,h}(z_1,\dots,z_h)=\Omega^\Knot_{g,h}(z_1,\dots,z_h)-\delta_{g,0}\delta_{h,2}\frac{\ud z_1\ud z_2}{(z_1-z_2)^2}.
\end{equation}
We also need the generating function for the invariants of knots in $\R P^3$
\begin{equation}
\Xi(\vec z)=\sum_{\lambda\in\Part} R_{\lambda}^{\Knot}(N_1,N_2)s_\lambda(\vec z)
\end{equation}
and the corresponding cumulants $\hat R_{\vec k}^{\Knot}(N_1,N_2)$. For reasons similar to the case of $\S^3$, $\log\Xi$ has an asymptotic expansion
\begin{equation}
\log \Xi=\sum_{g\ge 0}\sum_{\vec k\in\N^\N}\frac{1}{\mathfrak z_{\vec k}}\hbar'^{2g-2+\abs{\vec k}}  \hat R^{\Knot}_{\vec k,g} (S_1,S_2)p_{\vec k}(\vec z)
\end{equation}
where $\hat R^{\Knot}_{\vec k,g} (S_1,S_2)$ are power series with rational coefficients and $S_i=\hbar' N_i$ ($i=1,2)$. As an illustration, we reproduce some of the results obtained by \cite{Bouchard09a}.
\begin{align}
\hat R_{(1)}^{\Torus(1,0)}&=(S_1-S_2)+\frac 1 2 (S_1^2-S_2)^2\\
&\qquad\qquad\qquad +\frac{1}{24}(S_1-S_2)(4S_1^2+10S_1S_2+4S_2^2)+\O(S^4)\nonumber\\
\hat R_{(0,1)}^{\Torus(1,0)}&=(S_1+S_2)+2(S_1^2+S_2^2)\\
&\qquad\qquad\qquad +\frac13(S_1+S_2)(5S_1^2-2S_1S_2+5S_2^2)+\O(S^4)\label{rp3-T(1,0)}\nonumber\\
\hat R_{(2)}^{\Torus(1,0)}&=(S_1+S_2)+\frac 1 2 (3S_1^2+4S_1S_2+3S_2^2)\\
&\qquad\qquad\qquad +\frac 16(S_1+S_2)(7S_1^2+8S_1S_2+7S_2^2)+\O(S^4)\nonumber
\end{align}

We can now state our conjecture: for torus knots in $\R P^3$
\begin{equation}
\frac 1 2\int\omega_{g,h}^{\Torus(n,m)}=\sum_{\abs{\vec k}=h}\frac{1}{\mathfrak z_{\vec k}}\hat R^{\Torus(n,m)}_{\vec k,g} (S_1,S_2)p_{\vec k}(\vec z),\label{conjRP3}
\end{equation}
which is the analogous of \eqref{conjS3} and generalizes the case of $\Torus(1,0)$ proposed by Bouchard \textit{et al.} \cite{Bouchard09a}.

\subsection{Unknot}
We start with the simplest topological amplitude
\begin{equation}
\omega_{0,1}^{\Torus(1,0)}(X)=\log Y_{\Torus(1,0)}(X)\frac{\ud X}{X},
\end{equation}
where $Y_{\Torus(1,0)}(X)$ is obtained by solving $\mathcal E_{\Torus(1,0)}(X,Y)$ for $Y$,
\begin{equation}
Y_{\Torus(1,0)}(X)=\frac{(1+X+q_sX^2)+\sqrt{(1+X+q_sX^2)^2-4q_tX^2}}{2}.\label{yRP3}
\end{equation}
After integration, the disk amplitude reads
\begin{align}
\int\omega_{0,1}^{\Torus(1,0)}(X)&=\Big[(S_1-S_2)+\frac{1}{2} \left(S_1^2-S_2^2\right)\nonumber\\
&\qquad\qquad\qquad  +\frac{1}{12} \left(2 S_1^3+3 S_1^2 S_2-3 S_1 S_2^2-2 S_2^3\right) \nonumber\\
&\qquad\qquad\qquad+\frac{1}{24} \left(S_1^4+3 S_1^3 S_2-3 S_1 S_2^3-S_2^4\right)+\O(S^5)\Big]x\nonumber\\
&+\Big[(S_1+S_2) +2 \left(S_1^2+S_2^2\right)\\
&\qquad\qquad\qquad +\left(\frac{5 S_1^3}{3}+S_1^2 S_2+S_1 S_2^2+\frac{5 S_2^3}{3}\right) +\O(S^4)\Big]\frac{x^2}{2}\nonumber \\
&+ \O(x^3)\nonumber
\end{align}
and we see that it reproduces the CS knot invariants of the unknot \eqref{rp3-T(1,0)}.

The Bergman kernel is more complicated than for $\S^3$, because the curve is elliptic. By Akemann's formula \cite{Akemann96a}, $B(X_1,X_2)$ is expressed in terms of the ramification points $\{a_1,a_2,a_3,a_4\}$ as follows :
\begin{multline}
B(X_1,X_2)=\frac{E(k)}{K(k)}\frac{(a_1-a_3)(a_4-a_2)}{\sqrt{\prod_{i=1}^4 (X_1a_i-1)(X_2a_i-1)}}\\
+\frac{1}{4(X_1-X_2)^2}\Bigg[
\sqrt{\frac{(X_1a_1-1)(X_1a_2-1)(X_2a_3-1)(X_2a_4-1)}{(X_1a_3-1)(X_1a_4-1)(X_2a_1-1)(X_2a_2-1)}}\\
+\sqrt{\frac{(X_2a_1-1)(X_2a_2-1)(X_1a_3-1)(X_1a_4-1)}{(X_2a_3-1)(X_2a_4-1)(X_1a_1-1)(X_1a_2-1)}}+2\Bigg],\label{akemann}
\end{multline}
where $E(k)$ and $K(k)$ are elliptic functions of the first respectively second kind, and
\begin{equation}
k^2=\frac{(a_1-a_2)(a_3-a_4)}{(a_1-a_3)(a_2-a_4)}.
\end{equation}
This expression crucially depends on the ordering of the branch points. For our purpose, the correct ordering turns out to be \cite{Bouchard09a}
\begin{align}
a_1&=-\frac 1 2- \sqrt{q_t}-\frac 12\sqrt{(1+2\sqrt{q_t})^2-4q_s}\nonumber\\
a_2&=-\frac 1 2+ \sqrt{q_t}+\frac 12\sqrt{(1-2\sqrt{q_t})^2-4q_s}\\
a_3&=-\frac 1 2- \sqrt{q_t}+\frac 12\sqrt{(1+2\sqrt{q_t})^2-4q_s}\nonumber\\
a_4&=-\frac 1 2+ \sqrt{q_t}-\frac 12\sqrt{(1-2\sqrt{q_t})^2-4q_s}\nonumber.
\end{align}
This choice correspond to the orbifold point, and differs from the large-radius point considered in \cite{Marino08b} for instance.

Plugging in the ramification points in \eqref{akemann}, we obtain
\begin{align}
\frac 1 2\int\omega_{0,2}^{\Torus(1,0)}(X_1,X_2)&=\Big[(S_1+S_2)+\frac 1 2(3S_1^2+4S_1S_2+3S_2^2)\nonumber\\
&\qquad+\frac 1 6(7S_1^3+15S_1^2S_2+15S_1S_2^2+7S_2^3)\nonumber\\
&\qquad+\frac{1}{24}(15S_1^4+47S_1^3S_2+63 S_1^2S_2^2+47S_1S_2^3+15S_2^4)\nonumber\\
&\qquad+\O(S^5)\Big]x_1x_2\\
&+\Big[(S_1-S_2)+\frac 7 2(S_1^2-S_2^2)\nonumber\\
&\qquad+\frac 1{12}\big(62S_1^3+51S_1^2S_2-51 S_1S_2^2-62S_2^3\big)\nonumber\\
&\qquad+\frac{1}{24}(115S_1^4+201S_1^3S_2-201S_1S_2^3-115S_2^4)\nonumber\\
&\qquad+\O(S^5)\Big](x_1^2x_2+x_1x_2^2)+\O(\abs{\vec x}^4),\nonumber
\end{align}
where $X_1,X_2$ have been expressed in terms of $x_1,x_2$ by the mirror map \eqref{mirror}. Higher order amplitudes can be computed from the topological recursion \eqref{top-rec}.

Before extending our considerations to framed unknots and torus knots, one remark is in order. As one can observe in the above examples, our choice to consider $\Torus(1,1)$ instead of $\Torus(1,0)$ as the appropriate unknot in $\S^3$ is justified by the fact that
\begin{equation}
\frac 1 2\lim_{S_2\to 0}\omega_{0,1}^{\Torus(1,0)}(X;S_1,S_2)=\bar\omega_{0,1}^{\Torus(1,1)}(x;S_1),\label{limit(1,0)}
\end{equation}
which accounts for the relationship \eqref{rp3-qdim} between invariants of knots in $\S^3$ and $\R P^3$. Indeed, we shall show later that $\bar{\mathcal E}_{\Torus(1,1)}$ is obtained from $\mathcal E_{\Torus(1,0)}$ in the limit $S_2=0$.

\subsection{Framed Unknot}
The framed unknot in $\R P^3$ has been studied by Bouchard \textit{et al.} \cite{Bouchard09a}. We define $\eta_f$ as the solution $X=\eta_f(Z)$ of
\begin{equation}
Z=XY_{\Torus(1,0)}(X)^f,
\end{equation}
where $Y_{\Torus(1,0)}(X)$ is given by \eqref{yRP3}. Clearly, we obtain $\omega_{0,1}^{\Torus(1,f)}$ by re-para\-meterizing $\omega_{0,1}^{\Torus(1,0)}$,
\begin{equation}
\omega_{0,1}^{\Torus(1,f)}(Z)=\log Y_{\Torus(1,0)}\big(\eta_f(Z)\big)\frac{\ud Z}{Z}.
\end{equation}
We integrate $\omega_{0,1}^{\Torus(1,f)}(Z)$ with respect to $Z$
\begin{align}
\frac 1 2\int\omega_{0,1}^{\Torus(1,f)}(Z)&=\hat R_{(1);0}^{\Torus(1,f)}z+\hat R_{(0,1);0}^{\Torus(1,f)}\frac{z^2}{2}+\O(z^3)
\end{align}
and obtain full agreement with the knot invariants computed from CS theory.
\begin{align}
\hat R_{(1);0}^{\Torus(1,f)}&=(S_1-S_2) +\frac{1}{2} (1+2 f) (S_1^2-S_2^2)\nonumber\\
& +\frac{1}{12} (S_1-S_2) \big((2+6 f+6 f^2) S_1^2+(5+12 f+12 f^2) S_1 S_2+\dots\big)\nonumber \\
 &+\frac{1}{24} (1+2 f) (S_1^2-S_2^2) \big((1+2 f+2 f^2) S_1^2\\
 &\qquad\qquad\qquad\qquad+(3+4 f+4 f^2) S_1 S_2+\dots\big) +\O(S^5)\nonumber\\
\hat R_{(0,1);0}^{\Torus(1,f)}&=(S_1 + S_2)  + 2 (1 + 2 f) (S_1^2 + S_2^2)\nonumber\\
&  + \frac{1}{3} (S_1 + S_2) \big((5 + 18 f + 18 f^2) S1^2 -  2 (1 + 6 f + 6 f^2) S_1 S_2 + \dots\big)\nonumber \\
& + \frac{1}{12}\big((11 + 54 f + 96 f^2 + 64 f^3) S_1^4\\
&\qquad\qquad\qquad\qquad+  2 (9 + 34 f + 48 f^2 + 32 f^3) S_1^3 S_2 \nonumber\\
&\qquad\qquad\qquad\qquad\qquad\qquad
\qquad+ 6 (1 + 2 f) S_1^2 S_2^2 + \dots\big)+\O(S^5)\nonumber
\end{align}
The annulus amplitude is obtained by re-parameterization of the Bergman kernel \eqref{bergman},
\begin{align}
\omega_{0,2}^{\Torus(1,f)}(Z_1,Z_2)&=\Omega_{0,2}^{\Torus(1,0)}\big(\eta_f(Z_1),\eta_f(Z_2)\big)-\frac{\ud Z_1\ud Z_2}{(Z_1-Z_2)^2}\nonumber\\
&=\omega_{0,2}^{\Torus(1,0)}\big(\eta_f(Z_1),\eta_f(Z_2)\big)\\
&\qquad\qquad\qquad\qquad+\Bigg(\frac{\ud \eta_f(Z_1)\ud \eta_f(Z_2)}{\big(\eta_f(Z_1)-\eta_f(Z_2)\big)^2}-\frac{\ud Z_1\ud Z_2}{(Z_1-Z_2)^2}\Bigg).\nonumber
\end{align}
Integration yields
\begin{equation}
\frac 1 2\int\omega_{0,2}^{\Torus(1,f)}(Z_1,Z_2)=\hat R_{(2);0}^{\Torus(1,f)}z_1z_2+\O(\abs{\vec z}^3),
\end{equation}
where
\begin{align}
\hat R_{(2);0}^{\Torus(1,f)}&=(1 + 2 f) (S_1 + S_2)\nonumber\\
&\qquad  + \Big(\frac{3}{2} (1+ 2 f)^2S_1^2 +   2 (1 + 2 f + 2 f^2) S_1 S_2+\dots\Big) \\
&\qquad+ \frac{1}{6} (1 + 2 f) (S_1 + S_2) \big((7 + 24 f + 24 f^2) S_1^2\nonumber\\
&\qquad\qquad\qquad\qquad\qquad\qquad\quad +  8 S_1 S_2+\dots\big)+\O(S^4)\nonumber
\end{align}

\subsection{Torus Knots}
We now address the natural question whether the method of Brini \textit{et al.} can be applied to find a spectral curve for torus knots in $\R P^3$. We claim that the $SL(2,\ZZ)$-action on torus knots corresponds to the $SL(2,\ZZ)$ re-parameterization of the spectral curve \eqref{yRP3}. We consider the variable 
\begin{equation}
Z=X\Bigg[\frac{\big(1+(qX)+q_s (qX)^2\big)+\sqrt{\big(1+(qX)+q_s(qX)^2\big)-4q_t(qX)^2}}{2}\Bigg]^{m/n},
\end{equation}
where $X$ is multiplied by $q=\e^{\frac m n(S_1+S_2)}$, and the inversion
\begin{equation}
X=\eta_{\frac m n}(Z).
\end{equation}
We propose that the integer powers of $Z^n$ in the curve \eqref{yRP3} with fractional framing $m/n$ are identified with $\omega_{0,1}^{\Torus(n,m)}(Z^n)$,
\begin{equation}
\log Y_{\Torus(1,0)}\big(\eta_{\frac m n}(Z)\big)=\omega_{0,1}^{\Torus(n,m)}(Z^n)+\sum_{k\neq 0\,\mathrm{mod}\,n}\alpha_{0,1;k}^{\Torus(n,m)}Z^k.
\end{equation}

By comparing with the knot invariants computed from CS theory, we can verify conjecture \eqref{conjRP3} for several torus knots in $\R P^3$. Indeed, starting with $(2,m)$-torus knots we obtain the disk amplitude
\begin{multline*}
\int\log Y_{\Torus(1,0)}\big(\eta_{\frac m 2}(Z)\big)=\alpha_{0,1;1}^{\Torus(2,m)}z+\hat R_{(1);0}^{\Torus(2,m)}\frac{z^2}{2}+\alpha_{0,1;3}^{\Torus(2,m)}\frac{z^3}{3}\\
+\hat R_{(0,1);0}^{\Torus(2,m)}\frac{z^4}{4}+\O(z^5),
\end{multline*}
and check that the coefficients coincide with $\hat R_{(1);0}^{\Torus(2,m)}$ and $\hat R_{(0,1);0}^{\Torus(2,m)}$.
\begin{align}
\hat R_{(1);0}^{\Torus(2,m)}&=(S_1+S_2) +2 (1+m) \left(S_1^2+S_2^2\right)\nonumber\\
&\qquad+\frac{1}{6} (S_1+S_2)\Big(\left(10+18 m+9 m^2\right) S_1^2\nonumber\\
&\qquad\qquad-2 \left(2+6 m+3 m^2\right) S_1 S_2+\left(10+18 m+9 m^2\right) S_2^2\Big)\nonumber\\
&\qquad+\frac{1}{12} \Big(\left(11+27 m+24 m^2+8 m^3\right) S_1^4\\
&\qquad\qquad+2 \left(9+17 m+12 m^2+4 m^3\right) S_1^3 S_2\nonumber\\
&\qquad\qquad+6 (1+m) S_1^2 S_2^2+2 \left(9+17 m+12 m^2+4 m^3\right) S_1 S_2^3+\dots\Big)\nonumber\\
&\qquad +\O(S^5)\nonumber\\
\hat R_{(0,1);0}^{\Torus(2,m)}&=(S_1+S_2) +8 (1+m) \left(S_1^2+S_2^2\right)\nonumber\\
&\qquad+\frac{2}{3} (S_1+S_2) \big((34+66 m+33 m^2) S_1^2\\
&\qquad\qquad\qquad-2 (14+30 m+15 m^2) S_1 S_2+\dots\big)+\O(S^3)\nonumber
\end{align}
For $n=3$, the disk amplitude
\begin{equation}
\int\omega_{0,1}^{\Torus(1,0)}(\eta_{\frac m 3}(Z)\big)=\alpha_{0,1;1}^{\Torus(3,m)}z+\alpha_{0,1;2}^{\Torus(3,m)}\frac{z^2}{2}+\hat R_{(1);0}^{\Torus(3,m)}\frac{z^3}{3}+\O(z^7)
\end{equation}
agrees with the invariants computed from CS theory
\begin{multline}
\hat R_{(1);0}^{\Torus(3,m)}=(S_1-S_2)+\frac{3}{2} (3+2 m) \left(S_1^2-S_2^2\right)\\
+\frac{1}{12} (S_1-S_2) \big(2 \left(45+57 m+19 m^2\right) S_1^2\\
+\left(117+132 m+44 m^2\right) S_1 S_2+\dots\big)+\O(S^4)
\end{multline}

The conjecture can also be verified a step further for $(2,m)$-torus knots. The annulus amplitude
\begin{multline}
\int\Big(\Omega_{0,2}^{\Torus(1,0)}\big(\eta_{\frac m 2}(Z_1),\eta_{\frac m 2}(Z_2)\big)-\frac{\ud Z_1\ud Z_2}{(Z_1-Z_2)^2}\Big)=\alpha_{0,2;1,1}^{\Torus(2,m)}z_1z_2\\
+\alpha_{0,2;2,1}^{\Torus(2,m)}\frac{z_1^2z_2+z_2z_1^2}{2}+\hat R_{(2);0}^{\Torus(2,m)}\frac{z_1^2z_2^2}{4}+\O(\abs{\vec z}^5)
\end{multline}
matches
\begin{equation}
\hat R_{(2);0}^{\Torus(2,m)}=4 (1 + m) (S_1 + S_2) + 24 (1 + m)^2 (S_1^2 + S_2^2) +\O(S^3)
\end{equation}

As an additional evidence to support our proposal of a spectral curve for torus knots in $\R P^3$, we shall show that it reduces to the curve of Brini \textit{et al.} when $S_2\to 0$. A consistent picture emerges where the relation between $\S^3$ and $\R P^3$ knot invariants observed in CS theory has an analogous in the spectral curve description.
In particular, we claim that
\begin{equation}
\frac 1 2\lim_{S_2\to 0}\omega_{g,h}^{\Torus(n,m)}(X_1,\dots,X_h;S_1,S_2)=\bar\omega_{g,h}^{\Torus(n,2m+n)}(x_1,\dots,x_h;S_1),\label{limit(n,m)}
\end{equation}
which is the counterpart of \eqref{relS3RP3},
\begin{equation}
\lim_{S_2\to 0} \hat R_\lambda^{\Torus(n,m)}(S_1,S_2)=\hat H_\lambda^{\Torus(n,2m+n)}(S_1).
\end{equation}
In other words, the $SL(2,\ZZ)$-action on topological amplitudes commutes with the limit $S_2\to 0$, which can be summarized in the following diagram.
\begin{equation}
\xymatrix{\omega_{g,h}^{\Torus(1,0)}\ar[rrr]^{S_2\to 0} \ar[dd]_{SL(2,\ZZ)} & & &\bar\omega_{g,h}^{\Torus(1,1)}\ar[dd]^{SL(2,\ZZ)} \\
\\
\omega_{g,h}^{\Torus(n,m)}\ar[rrr]^{S_2\to 0} & & & \bar\omega_{g,h}^{\Torus(n,2n+m)}}
\end{equation}
Let us finally prove \eqref{limit(n,m)} for the disk amplitude. Starting from 
\begin{equation}
Y_{\Torus(1,1)}(X)=\frac{(1+X+q_sX^2)+\sqrt{(1+X+q_sX^2)^2-4q_tX^2}}{2},
\end{equation}
we change the variables to $x=\sqrt{q_s} X=\frac{1}{q_1q_2}X$ and introduce $\alpha,\beta$ as
\begin{equation}
\alpha-\beta=\frac{2}{\sqrt{q_s}}\qquad\text{and}\qquad\alpha+\beta=4\sqrt{\frac{q_t}{q_s}}.
\end{equation}
We obtain
\begin{align}
Y_{\Torus(1,1)}(x/\sqrt{q_s})&=\frac 1 2\Big(1+\frac{\alpha-\beta}{2}x+x^2\Big)\nonumber\\
&\qquad\qquad+\frac 1 2\sqrt{\Big(1+\frac{\alpha-\beta}{2}x+x^2\Big)^2-\Big(\frac{\alpha+\beta}{2}\Big)^2x^2}\\
&=\Bigg[\frac{\sqrt{1+\alpha x+x^2}+\sqrt{1-\beta x+x^2}}{2}\Bigg]^2.\nonumber
\end{align}
In the limit $S_2\to 0$, the spectral curve $\mathcal E_{\Torus(1,0)}$ reduces to $\bar{\mathcal E}_{\Torus(1,1)}$,
\begin{align}
\lim_{S_2\to 0}Y_{\Torus(1,0)}(X)&=\Big[\frac{\sqrt{1+2 x+x^2}+\sqrt{1+(2-4q) x+x^2}}{2}\Big]^2\\
&=\Big[\frac{(1+x)+\sqrt{(1+x)^2-4v^2 x}}{2}\Big]^2=y_{\Torus(1,1)}(x)^2\nonumber
\end{align}
In turn, $ Z = X Y(X)^f$ reduces to $z=x y(x)^{2f}$, hence a simple comparison shows that
\begin{equation}
\lim_{S_2\to 0} \frac{1}{q_1q_2}\eta_f(Z)=\bar\eta_{2f}(z).
\end{equation}
Since $\omega_{0,1}^{\Torus(n,m)}$ is obtained by re-parameterization of $\omega_{0,1}^{\Torus(1,0)}$, we deduce \eqref{limit(n,m)}.

\section{Conclusion}
We have studied invariants of torus knots in lens spaces from the point of view of quantum Chern-Simons theory. Using the knot operator formalism developed by Labastida and collaborators \cite{Labastida89a,Labastida91a}, we have derived a generalization of the Rosso-Jones formula \cite{Rosso93a} for knots in lens spaces. Using the matrix model representation found by Bouchard \textit{et al.} \cite{Bouchard09a} for the unknot in $L(p,1)$, our formula provides computable expressions for torus knots in $L(p,1)$. For the specific case of $L(2,1)\approx\R P^3$, we have proposed an algebraic curve that describes torus knots in the B-model topological string theory.

Along the same lines, one could in principle study generalizations to lens spaces of the colored Kauffman polynomial. Invariants of the rational unknot can probably be written as a matrix integral and the generalized Rosso-Jones formula gives invariants of torus knots. In a certain limit, the invariants should correspond to the generalization of the Kauffman polynomial developed by Kalfagianni \cite{Kalfagianni10a}. The question whether there exist an algebraic curve description for Kauffman invariants in $L(p,q)$ is completely open. Already in $\S^3$, there is no such description for the Kauffman polynomial, and the enumerative description \cite{Marino09c} is much more complicated than for the HOMFLY case.

For HOMFLY invariants at least, the conjecture which relates invariants of torus knots to re-parameterizations of the spectral curve \cite{Halmagyi09a} undoubtedly extends to $L(p,1)$ with $p>2$. On the Chern-Simons side, the invariants can be computed, at least perturbatively, as a power series in $p+1$ variables. On the topological string theory side, the topological recursion is more difficult to implement if no hyperelliptic embedding of the spectral curve is known. More importantly, the comparison involves a very complicated mirror map.

The question of $L(p,q)$ with $q\neq 1$ is also interesting from the point of view of topological string theory. The Gopakumar-Vafa duality, in its current formulation, does not hold for $q\neq 1$. Related to this, it might be fruitful to study the full partition function $Z\big(L(p,q)\big)$ and the corresponding knot invariants. Equivalently, this amounts to studying instanton effects in Chern-Simons matrix models.


\end{document}